\documentclass[a4paper,11pt]{article}
\usepackage{amsmath}
\usepackage{amssymb}
\usepackage{indentfirst}
\usepackage[english]{babel}
\usepackage{graphicx} 
\usepackage{subcaption}
\usepackage[dvipsnames,svgnames,table]{xcolor}
\usepackage{epstopdf}
\usepackage{mathrsfs}
\usepackage{ulem}
\usepackage{enumerate}
\usepackage{enumitem}
\usepackage{amsthm}
\usepackage{cite}
\usepackage{float}
\usepackage{xcolor,soul}
\usepackage{booktabs}
\usepackage{cleveref}

\newcommand{\be}{\begin{equation}}
\newcommand{\ee}{\end{equation}}
\newcommand{\bea}{\begin{eqnarray}}
\newcommand{\eea}{\end{eqnarray}}

\newcommand{\eal}{{\it et al.} }

\newcommand{\volume}{{\ooalign{\hfil$V$\hfil\cr\kern0.08em--\hfil\cr}}}

\DeclareMathAlphabet\mathbfcal{OMS}{cmsy}{b}{n}

\def\pmb#1{\setbox0=\hbox{#1}
	\kern-.025em\copy0\kern-\wd0
	\kern.05em\copy0\kern-\wd0
	\kern-.0125em\raise.0433em\box0 } % hand made bold characters

\def\bv{\pmb {v}}

\def\bbe{\pmb {e}}

\def\bV{\pmb {V}}

\def\rd{{\mathrm{d}}}

\title{Host-to-Host Airborne Transmission As a Multiphase Flow Problem For Science-Based Social Distance Guidelines}

\author{
	S. Balachandar, 
	\thanks{Corresponding Author, Professor, University of Florida, Gainesville, FL, USA, bala1s@ufl.edu} \,
	S. Zaleski,
	\thanks{Professor, Sorbonne Universit\'e,  Institut Jean Le Rond d'Alembert, Paris, France, stephane.zaleski@sorbonne-universite.fr; CNRS, Institut Jean Le Rond d'Alembert, Paris, France; Senior Member, Institut Universitaire de France (IUF), Paris, France} \,
	A. Soldati,
	\thanks{Professor, TU Wien, Vienna, Austria, \& University of Udine, Udine, Italy alfredo.soldati@tuwien.ac.at} \,
	G. Ahmadi,
	\thanks{Professor, Clarkson University, Potsdam, NY, USA, ahmadi@clarkson.edu} \,
	L. Bourouiba
	\thanks{Professor, Massachusetts Institute of Technology, Boston, MA, USA, lbouro@mit.edu} \,
	\\
	}

\date{}

\begin{document}
	\maketitle
	
\begin{abstract}

%\Lnote{Notes are left in text from LB in red. Direct edits of context and some points about evaporation, in light of our current data collection are made directly without color when minor, and with color when major, so discussion of such points can already incorporates our early insights form lab ongoing at this time on evaporation and dispersal in particular. Note that I changed expectoration to exhalation (and derivatives of the word), as expectoration is a bit more of a specific type of exhalation in medical jargon.)}

COVID-19 pandemic has strikingly demonstrated how important it is to develop fundamental knowledge related to generation, transport and inhalation of pathogen-laden droplets and their subsequent possible fate as airborne particles, or aerosols, in the context of human to human transmission. It is also increasingly clear that airborne transmission is an important contributor to rapid spreading of the disease. In this paper, we discuss the processes of droplet generation by exhalation, their potential transformation into airborne particles by  evaporation, transport over long distances by the exhaled puff and by ambient air turbulence, and final inhalation by the receiving host as interconnected multiphase flow processes. A simple model for the time evolution of droplet/aerosol concentration is presented based on a theoretical analysis of the relevant physical processes. The modeling framework along with detailed experiments and simulations can be used to study a wide variety of scenarios involving breathing, talking, coughing and sneezing and in a number of environmental conditions, as humid or dry atmosphere, confined or open environment. {Although a number of questions remain open on the physics of evaporation and coupling with persistence of the virus, it is clear that with a more reliable understanding of the underlying flow physics of virus transmission one can set the foundation for an improved methodology in designing case-specific social distancing and infection control guidelines.} 

\end{abstract}

\section{Introduction}	\label{sec:1}

{The COVID-19 pandemic has made clear the fundamental role of airborne droplets and aerosols as potential virus carriers. The importance of studying the fluid dynamics of exhalations, starting from the formation of droplets in the respiratory tracts to their evolution and transport as a turbulent cloud, can now be recognized as the key step towards understanding SARS-CoV-2 transmission. Respiratory }droplets are formed and emitted at high speed during a sneeze or cough \cite{scharfman2016visualization}, and at lower speed while talking or breathing. The virus-laden droplets are then initially transported as part of the coherent gas puff of buoyant fluid ejected by the infected host \cite{Bourouiba2014}. The very large drops of $O(mm)$ in size, which are visible to naked eye, are minimally affected by the puff. They travel semi-ballistically with only minimal drag adjustment, but rapidly falling down due to gravitational pull. They can exit the puff either by overshooting or by falling out of the puff at the early stage of emission (Fig. \ref{fig0}). Smaller droplets ($\lesssim O(100 \, \mu m)$) that remain suspended within the puff are advected forward. {As the suspended droplets steadily evaporate within the cloud, the virus takes the form of potentially inhalable droplets, or droplet residues when the evaporation of water is complete. Meanwhile, the velocity of the turbulent puff continues to decay both due to entrainment and drag. Once the puff slows down sufficiently, and its coherence is lost, the eventual spreading of the virus-laden droplet residue becomes dependent on the ambient air currents and turbulence. }

{The isolated respiratory droplet emission framework was introduced by Wells \cite{wells1955airborne} in the 1930s and remains the framework used for guidelines by public health agencies, such as the WHO, CDC and others. However, it does not  consider the role of the turbulent gas puff within which the droplets are embedded.} Regardless of their size and their initial velocity, the ejected droplets are subject to both gravitational settling and evaporation \cite{Bourouiba2014}. Although droplets of all sizes undergo continuous settling, droplets with settling speed smaller than the fluctuating velocity of the surrounding puff can remain trapped longer within the puff (Fig.~\ref{fig1}). Furthermore, the water content of the droplets continuously decreases due to evaporation. When conditions are appropriate for near complete evaporation, the ejected droplets quickly become droplet residues of non-volatile biological material. The settling velocity of these droplet residues is sufficiently small that they can remain trapped as a cloud and get advected by ambient air currents and dispersed by ambient turbulence. Based on the above discussion, we introduce the following terminology that will be consistently used in this paper:
\begin{itemize}
	\item {\it {Puff:}} Hot, moist air exhaled during breathing, talking, coughing or sneezing, which remains coherent and moves forward during early times after exhalation
	\item {\it {Cloud:}} The distribution of ejected droplets that remain suspended even after the puff has lost its coherence. The cloud is advected by the air currents and is dispersed by ambient turbulence
	\item {\it {Exited droplets:}} droplets that have either overshot the puff/cloud or settled down due to gravity
	\item {\it {Airborne (evaporating) droplets:}} droplets which have not completed evaporation and retained within the puff/cloud
	\item {\it {(Airborne) droplet residues:}} droplets that remain airborne within the puff/cloud and that have fully evaporated, which will also be termed {\it {aerosols}}.
\end{itemize}

Host-to-host transmission of virus-laden droplets and droplet residues occurs generally through direct and indirect routes \cite{Atkinson, Pan, Bourouiba}. {{ The direct route of transmission involves the larger droplets that may ballistically reach the recipient's mucosa. This route is currently thought to  involve either the airborne route or drops that have settled on surfaces. The settled drops remain infectious, to be later picked up by the recipient, and are generally thought to be localized to the vicinity or at close range of the original infectious emitter. With increased awareness and modified physical distancing norms it is possible to minimize the spreading of the virus by such direct route. }

%The indirect route of transmission is through the smaller droplets and aerosols released to the surrounding by the infected individual that can stay afloat. As larger droplets settle quickly to the ground by gravity, direct contact transmission occurs within a short distance of less than 6 feet, whereas in airborne indirect transmission, spreading of the virus can be over a longer distance (10-100 feet) and a longer period of time (see Fig. \ref{fig1}) and therefore the indirect transmission route is of great concern \cite{Shiu}. Recent studies show that airborne transmission is more efficient than direct-contact transmission, accounting for a large fraction of all transmission \cite{Smieszek}. 

\begin{figure}[hbt!]		   	
	\begin{center}
		\includegraphics[width=0.49\linewidth]{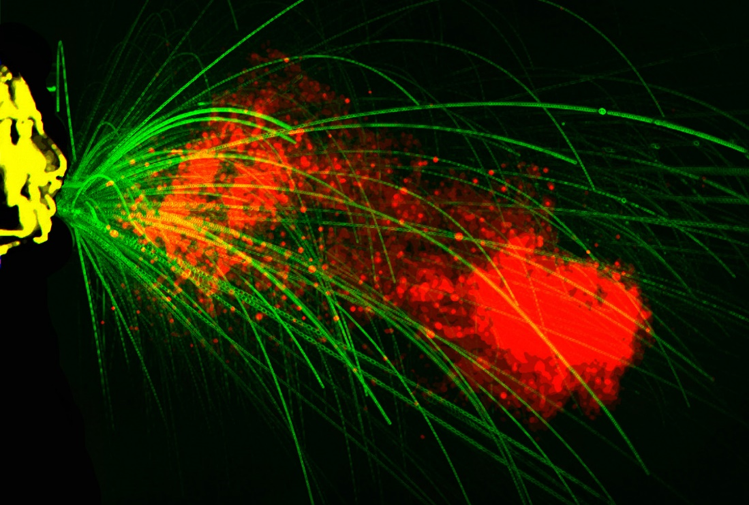}
	\end{center}   
	\caption{Image reproduction showing the semi-ballistic largest drops, visible to the naked eye, and on the order of mm, which can overshoot the puff at its early stage of emission \cite{BourouibaHHMI, BourouibaNEJM}. The puff continues to propagate and entrain ambient air as it moves forward, carrying its payload  of continuum of drops \cite{Bourouiba2014}, over distances up to 8 meters for violent exhalations such as sneezes \cite{jama}.}
	\label{fig0}
\end{figure}

{The indirect route of transmission is one that does not necessarily involve a direct or close interaction between the infectious individual and the recipient or for the two to be synchronously present in the same contaminated space at the same time.  Thus, the indirect route involves respiratory droplets and fully-evaporated droplet residues that are released to the surrounding by the infected individual, which remain airborne as the cloud carries them over longer distances. The settling speeds of the airborne droplets and droplet residues are so small, that they remain afloat for longer times, while being carried by the background turbulent airflow over distances that can span the entire room or even multiple rooms within the building ($O(10-100)$ feet). } A schematic of the two routes of transmission is shown in Fig.~\ref{fig1} and in this paper we will focus on the indirect airborne transmission. 
%The infected individual, via exhalations from sneezing, coughing, talking or breathing produces a large number of droplets of varying sizes, the size distribution of which depends on a number of factors, but in almost all cases {the ejected droplet size extends over the range O(1-100 $\mu$m).} 
%\Lnote{Here and other locations, it would be beneficial, and what I have been aiming to do over the past years, is to stay away from the dichotomy of large vs small drops so we can, as a community, reframe the discussion between air vs. surface, with timescales, and spatial scales, rather than the artificial cutoff of drop sizes - I tried to change this throughout for consistency - for your consideration}. 

{Another factor of great importance is the possibility of updraft in the region of contamination, due to buoyancy of the virus-laden warm ejected air-mass. These slight updrafts can keep the virus-laden droplets suspended in the air and enhance the inhalability of airborne droplets and droplet residues by recipients who are located farther way.} The advection of airborne droplets and residues by the puff and subsequently as a cloud may represent transmission risk for times and distances much longer than otherwise previously estimated, and this is a cause of great concern~\cite{Shiu, Smieszek}. {Note that if we ignore the motion of the puff of air carrying the droplets, as in the analysis of Wells, the airborne droplets and residues would be subjected to such high drag that they could not propagate more than a few $cm$ away from the exhaler, even under conditions of fast ejections, such as in a sneeze. This illustrates the importance of incorporating the correct multiphase flow physics in the modeling of respiratory emissions \cite{ARFM2020}, which we shall discuss further here.}

It has been recently reported that COVID-19 virus lives in droplets and aerosols for many hours in laboratory experiments \cite{Doremalen}. At the receiving end, an increased concentration of virus-laden airborne droplets and residues near the breathing zone increases the probability of them settling on the body or more importantly being inhaled. Depending on its material and sealing properties, the use of a mask by the infected host {can help} reduce the number of virus-laden droplets ejected into the air, and in a less effective way, the use of a mask or other protective devices by the receiving host may reduce the probability of inhalation of the virus-laden airborne droplets and residues.

The above description provides a clear sketch of the sequence of processes by which the virus is transferred host-to-host. This simplistic scenario, though pictorially evocative, is tremendously insufficient to provide science-based social distancing guidelines and recommendations. There is substantial variability (i) in the quantity and quality of contaminated droplets and aerosols generated by an infected person, (ii) in the manner by which the contaminated droplets and droplet residues remain afloat over longer distances and time, (iii) in the possibility of the contaminant being inhaled by a recipient and (iv) in the effectiveness of masks and other protection devices. Violent {exhalations}, such as sneezing and coughing, yield many more virus-laden droplets and aerosols than breathing and talking \cite{Bourouiba2014, Memarzadeh}. All coughing and sneezing events are not alike - the formation of droplets by break up of mucus and saliva varies substantially between individuals. Significant variation in initial droplet size and velocity distribution has been reported in \cite{Han, Bourouiba2014, Asadi, Pan}. {The measured droplet size distribution, particularly for transient biological emissions such as respiratory exhalations, also depends on ambient temperature and humidity and on the methodology and instrumentation used to characterize the size distribution \cite{Atkinson, Memarzadeh, ARFM2020}.} 
%After the initial exhalation of polydisperse droplets, their advection and dispersion is in the form of a turbulent multiphase puff.
Furthermore, it is of importance to consider the volume of air, {and the pathogen load}, being inhaled during breathing by the receiving host. Thus, there is great variability in how much of the virus-laden aerosols reach from the infected host to the receiving host. 

CDC guideline of social distancing of 2 meters (6 feet) is based on the disease transmission theory originally developed in 1930s and later improved by others {\cite{Wells, xie2007, Ox}. The current recommendation of 6 feet as the safe distance can be improved in several ways: (i) by accurately accounting for the distance traveled by the puff and the droplets contained within it, while some continuously settling out of the puff, (ii) by accurately evaluating the evaporation of droplets and the subsequent advection and dispersal of droplet residues as a cloud \cite{Bahl},} (iii) by incorporating the effect of adverse flow conditions that prevail under confined indoor environment including elevators, aircraft cabins, and public transit, or favorable conditions of open space with good breeze or cross ventilation, and (iv) by correctly assessing the effectiveness of masks and other protective devices {\cite{Cooper}. Thus, mechanistic, evidence-based  understanding of exhalation and dispersal of expelled respiratory droplets, and their subsequent fate as droplet residues in varying scenarios and environments is important.} We must therefore revisit the safety guidelines and update them to modern understanding. In particular, a multi-layered guideline that differentiates between crowed class rooms, auditoriums, buses, elevators and aircraft cabins from open outdoor cafes is desired. Only through reliable understanding of the underlying flow physics of virus transmission one can arrive at such nuanced guidance in designing case-specific social distancing guidelines.

The object of the paper is to present a coherent analytic and quantitative description of droplets generation, transport, conversion to droplet residues, and eventual inhalation. We will examine the available quantitative relationships that describe the above processes and adapt them to the present problem. The key outcomes that we desire are (i) A simple universal description of the initial droplet size spectrum generated by sneezing, coughing, talking and breathing activities. {Such a description must recognize the current limitations of measurements of droplet size distribution under highly transient conditions of respiratory events. (ii) A  first-order mathematical framework that describes the evolution of the cloud of respiratory droplets and their conversion to droplet residues, as a function of time,  and (iii) A simple description of the inhalability of the aerosols  along with corresponding evaluation of the effectiveness of different masks based on existing data reported to date.}
%for some known material filtration efficacy.} 
The physical picture and the quantitative results to be presented can then be used to study a statistical sample of different scenarios and derive case-specific guidelines. We anticipate the present paper to spawn future research in the context of host-to-host airborne transmission. 

After presenting the mathematical framework in section 2, the three different stages of transmission, namely droplet generation, transport and inhalation will be independently analyzed in sections 3, 4 and 5. These sections will consider the evolution of the puff of exhaled air and the droplets contained within. Section 6 will put together the different models of the puff and droplet evolution described in the previous sections, underline their simplifications, and demonstrate their ability to make useful prediction. Finally, conclusions and future perspectives are offer in section 7.

\begin{figure}[hbt!]		   	
	\begin{center}
		\includegraphics[width=0.6\linewidth]{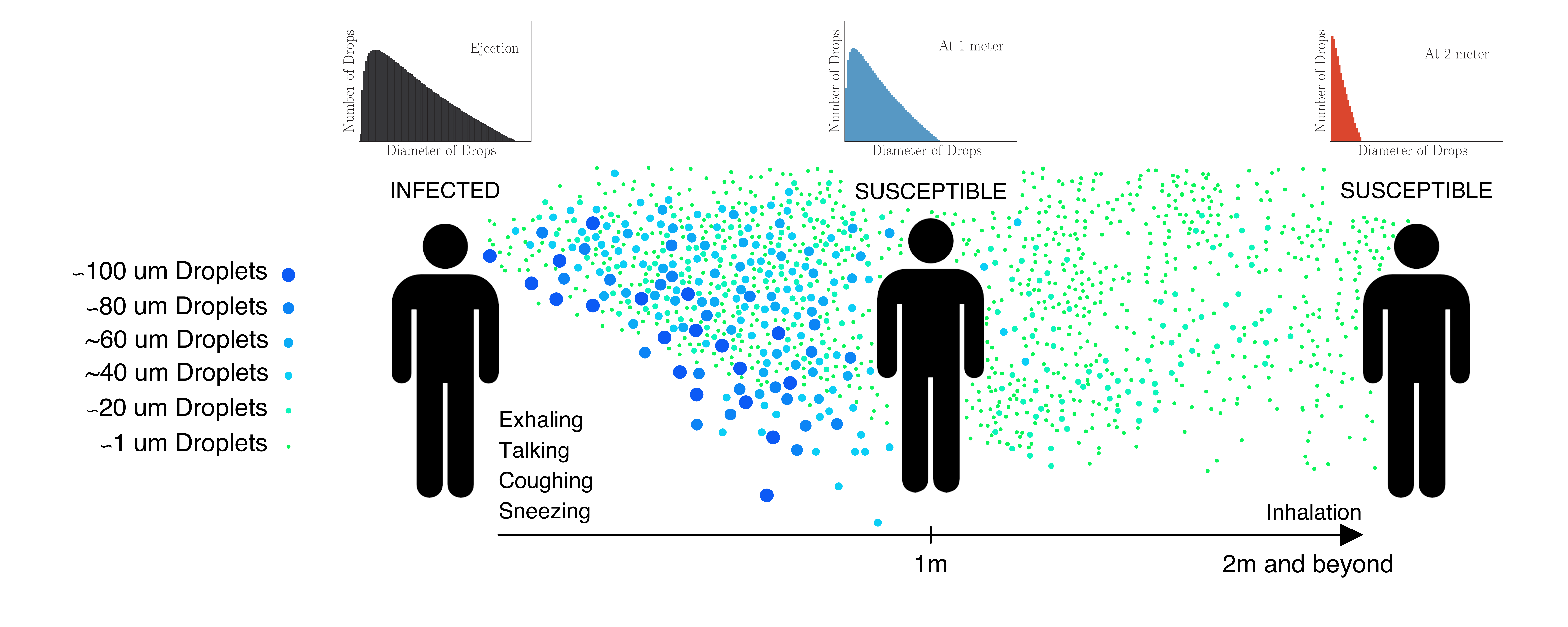}
	\end{center}   
	\caption{The two dominant transmission routes (a) direct transmission route through ballistic larger droplets (b) indirect airborne transmission route by smaller airborne droplets and droplet residues. A schematic representation of size distribution at the infected source host, at an intermediate distance and at a receiving host located farther away is also shown.}
	\label{fig1}
\end{figure}

\section{Problem Description and Mathematical Framework}
%\Lnote{Maybe referring reader to sub-sections with more upcoming details when they are covered later?}
We wish to describe the three main stages involved in the host-to-host transmission of the virus: droplet generation during exhalation, airborne transport, and inhalation by the receiving host. In the {{generation stage}}, virus-laden drops are generated {throughout the respiratory tract by the exhalation air flow which carries them through the upper airway toward the mouth where}  they are ejected along with {the turbulent puff of air from the lungs. The ejected  puff of air} can be characterized with the following four parameters: the volume $Q_{pe}$, the momentum $M_{pe}$, and the buoyancy $B_{pe}$ of the ejected puff, along with the angle $\theta_{e}$ to the horizontal at which the puff is initially ejected. The initial momentum and buoyancy of the puff are given by %\Lnote{there was a missing $\rho_{pe}$ added in the momentum $M_{pe}$, but may need to be homogenized for notation later in text?}} 
$M_{pe} = \rho_{{pe}} Q_{pe} v_{pe}$ and $B_{pe} = (\rho_{a}-\rho_{pe}) Q_{pe} g$, 
%\Lnote{Buoyancy is defined to be negative if cloud is lighter than ambient - typo of sign,  reverse here and subsequently?} 
where $v_{pe}$ is the initial velocity of ejected puff, $\rho_{pe}$ and $\rho_a$ are the {initial} density of {the puff} and the ambient, respectively, and $g$ is the gravitational acceleration. The ejected droplets are characterized by their total number ${\cal N}_{e}$, size distribution $N_e(d)$, droplet velocity distribution $V_{de}(d)$ and droplet temperature distribution $T_{de}(d)$, where $d$ is the diameter of the droplet. To simplify the theoretical formulation, here we assume the velocity and temperature of the ejected droplets to depend only on the diameter and show no other variation. As we shall see in section 4, this assumption is not very restrictive, since the velocity and temperature of the droplets that remain within the puff very quickly adjust to those of the puff. Both the ejected puff of air and the detailed distribution of droplets depend on the nature of the exhalation event (i.e., breathing, talking, coughing or sneezing), and also on the individual. 

This is followed by the {{transport stage}}, where the initially ejected puff of air and droplets are transported away from the source. The volume of the puff of air increases due to entrainment of ambient air. The puff velocity decreases due to both entrainment as well as drag. Since the temperature {and moisture content} of the ejected puff of air is typically higher than the ambient, the puff is also subjected to a vertical buoyancy force, which alters its trajectory from a rectilinear motion. The exhaled puff is turbulent and both the turbulent velocity fluctuations within the puff and the mean forward velocity of the puff decay over time.  

The time evolution of the puff during the transport stage can then be characterized by the following quantities: the volume $Q_p(t)$, the momentum $M_p(t)$, buoyancy $B_p(t)$ of the ejected puff, {and $\rho_p(t)$ is the density of air within the puff which changes over time due to entrainment and evaporation}. The trajectory of the puff is defined in term of the distance traveled $s(t)$ and the angle to the horizontal $\theta(t)$ of its current trajectory. Following the work of Bourouiba {\it et al.} \cite{Bourouiba2014} we have chosen to describe the puff trajectory in terms of $s(t)$ and $\theta(t)$. This information can be converted to horizontal and vertical positions of the {centroid of the} puff as a function time. If we ignore the effects of thermal diffusion {and ambient stratification} between the puff and the surrounding air, then the buoyancy of the puff remains a constant as $B_p(t) = B_{pe}$.  Furthermore, as will be seen below, the buoyancy effects are quite weak in the early stages when the puff remains coherent and thus the puff to good approximation can be taken to travel along a straight line path, as long as other external flow effects are unimportant.

To characterize the time evolution of the virus-laden droplets during the transport stage we distinguish the  droplets that remain within the puff, whose diameter is less than a cutoff (i.e., $d < d_{exit}$), from the  droplets (i.e., $d > d_{exit}$) that escape out of the puff. {As will be discussed subsequently in \S \ref{sec4}}, the {\it {cutoff droplet size}} $d_{exit}$ decreases with time - at very early times it decreases as $t^{-1/4}$ when the ejection behaves as a jet, then decays as $t^{-1/2}$ when evolving as a puff and at later times the cutoff diameter decreases as $t^{-3/8}$. %(these scaling relations will be established later). 
Thus, the total number of droplets that remain within the puff can be estimated as ${\cal N}(t) = \int_0^{d_{exit}} N(d,t) \, \rd d$. However, the size distribution of droplets at any later time, denoted as $N(d,t)$, is not the same as that at ejection. Due to evaporation, size distribution shifts to smaller diameters over time. We introduce the mapping $\mathcal{D} (d_{e},t)$, which gives the current diameter of a droplet initially ejected as a droplet of diameter $d_e$. Then, assuming well-mixed condition within the puff, the airborne droplet and residue concentration (number per volume) distribution can be expressed as
\be
\phi(d,t) = \dfrac{N(d,t)}{Q_p(t)} =  \dfrac{1}{Q_p(t)} N_e(\mathcal{D}^{-1}(d,t)) \quad \mathrm{for} \quad 0 \le d \le d_{exit} \, ,
\ee
where the inverse mapping $\mathcal{D}^{-1}$ gives the original ejected diameter of a droplet whose current size is $d$. The prefactor $1/Q_p(t)$ accounts for the decrease in concentration due to the enlargement of the puff over time. In this model the airborne droplets and residue that remain within the coherent puff are assumed to be in equilibrium with the turbulent flow within the puff. Under this assumption, the velocity $V_d(d,t)$ and temperature $T_d(d,t)$ of the droplets can be estimated with the equilibrium Eulerian approximation \cite{bala-eea,bala-thermal-eea}.

When the puff's mean and fluctuating velocities fall below those of the ambient, the puff can be taken to loose its coherence. Thus, the puff remains coherent and travels farther in a confined relatively quiescent environment, such as an elevator, class room or aircraft cabin, than in an open outdoor environment with cross-wind or in a room with {strong ventilation}. We define a {\it {transition time}} $t_{tr}$, below which the puff is taken to be coherent and the above described puff-based transport model applies. For $t > t_{tr}$, we take the aerosol transport and dilution to be dominated by ambient turbulent dispersion. Accordingly, this late-time behavior of total number of airborne droplets and residue and their number density distribution are given by theory of turbulent dispersion. It should be noted that the value of transition time will depend on both the puff properties as well as the level of ambient turbulence (see section \ref{sec4.4}).

We now consider the final {{inhalation stage}}. Depending on the location of the recipient host relative to that of the infected host, the recipient may be subjected to either the puff {that still remains coherent, carrying a relatively high concentration of virus-laden droplets or residues}, or to the {more} dilute dispersion of droplet residues, or aerosols. These factors determine the number and size distribution of virus-laden airborne droplets and residue the recipient host will be subjected to. The inhalation cycle of the recipient, along with the use of mask and other protective devices will then dictate the aerosols that reach sensitive areas of {the respiratory tract} where infection can occur. 
%The entire dynamics of transport and deposition is controlled by the evolution of droplet and aerosol size and velocity, which are ever changing characteristics during the entire process. 
Following the above outlined mathematical framework we will now consider the three stages of generation, transport and inhalation.

\section{Ejection Stage} \label{sec3}

\renewcommand\Re{{\rm Re}\,}
\newcommand\We{{\rm We}\,}

Knowing the droplet sizes, velocities and ejection angles resulting
from an exhalation is the key first step in the development
of a predictive ability for droplet dispersion and evolution. {Respiratory} droplet
size distributions have been the object of a large
number of studies, as reviewed in \cite{Gralton:2011kq},
and among them, those of Duguid \cite{Duguid:1946dw} and Loudon \&
Roberts \cite{loudon1967relation} have received particular scrutiny as
a basis for studies of disease transmission by Nicas, Nazaroff \&
Hubbard \cite{Nicas:2017eo}. {There are substantial differences in the methodologies used for quantification of respiratory emission sprays. Few studies have used common instrumentation that have enough overlap to reconstruct the full distribution of sizes. For example, there are important gaps in reporting the total volume or duration of air sampling, in addition there are issues in reporting the effective evaporation rates used to back-compute the initial distribution and in the documentation of assumptions about optical or shape properties of the droplets being sampled. In addition, sensitivity analyses are often missing regarding the role of orientation or calibration of sensing instruments with respect to highly variable emissions from human subjects. Finally, regarding direct high-speed imaging methods \cite{Bahlexp, scharfman2016visualization}, the tools for precise quantification of complex unsteady fragmentation and atomization processes are only now being developed \cite{ARFM2020, Wang:2018du, Wang:2020}}.
%and necessary adaption of measurement methods to improve accuracy and modeling of such unsteady flow distributions. }  
There are far fewer studies on the velocities and angles of the droplets produced by atomizing flows. 
%{Nevertheless,  we shall see below, how capturing a portion of reported distributions can enable substantial progress in the assessment of the droplet dynamics. }

\subsection{Droplet sizes} \label{sec3.1}

The studies of Duguid and Loudon \& Roberts have been performed by
allowing the exhaled  droplets to impact various sheets or slides,
with different procedures being used for droplets smaller than
20 $\mu m$. The size of the stains on the sheets was observed and the
original droplet size was inferred from the size of the stains. To
account for the difference between the droplet and the stain sizes an arbitrary factor
is applied and droplets smaller than 10 or 20 microns are processed
differently than larger droplets.  The whole process makes the
determination of the number of droplets smaller than 10 microns less
reliable. The data are replotted in Fig. \ref{replot}.

Many authors have attempted to fit the data with a log-normal probability distribution function. In that case, the number of droplets
between diameter $d$ and $d+\rd d$ is $N_e(d) \, \rd d$, and the frequency of ejected droplet size distribution is given by 
\begin{equation}
\mathrm{Log-Normal \; Distribution:} \quad  N_e(d) = \frac{B}{d} \exp\left[ - \frac{ (\ln d - \hat \mu)^2}{2 \hat \sigma^2} \right], \label{ln}
\end{equation}
where $\rd d$ is a relatively small diameter increment or bin width, $B$ is a normalization constant, $\hat \mu$ is the expected value of $\ln d$, also called the {\it geometric mean} and
$\hat \sigma$ is the standard deviation of  $\ln d$, also called the {\it geometric standard deviation} (GSD).

On the other hand, there have been also numerous studies of the fragmentation of liquid masses in various physical configurations other than the exhalation of mucosalivary fluid \cite{Villermaux2007, gorokhovski2008modeling, scharfman2016visualization, Wang:2018PRL}. These configurations include spray formation on wave crests \cite{Veron:2015hy}, droplet impacts on solids and liquids \cite{mundo1995droplet}, wave impacts on vertical or finite walls/surfaces \cite{watanabe2015transverse,Wang:2018du,Lejeune}, and jet atomization \cite{Eggers08}. These studies reveal a number of qualitative similarities between the various processes, which can be best described as a cascade of events.
Those events include a primary
instability of sheared layers in high speed air flows \cite{Fuster2013}, and then the non linear growth of the perturbation into thin liquid sheets. The sheets themselves may be destabilized by two routes, one involving the formation of Taylor-Culick end rims \cite{Taylor59c,Culick60}, and their subsequent deformation into detaching droplets \cite{Wang:2018du}. The other route to the formation of droplets is the formation of holes in the thin sheets \cite{opfer14,lhuissier2013effervescent,scharfman2016visualization}. The holes then expand and form free hanging ligaments which fragment into droplets through
the Rayleigh-Plateau instability \cite{Eggers08}.

Considering the apparent universality of the process, one may infer that a universal distribution of droplet sizes may exist. Indeed, the log-normal distribution has often been fitted to experimental \cite{Marty:2015vs} and numerical data on jet formation
\cite{Herrmann:2011fq,Ling17}, for droplet impacts on solid surfaces \cite{wu2003prediction}, and for wave impacts on solid walls \cite{wu2003prediction}. The log-normal distribution is frequently suggested for exhalations
\cite{Nicas:2017eo,wells1955airborne}. The fit of the numerical results of
\cite{Ling17}   is shown in Fig. \ref{ling2}. However, this apparent universality of the log-normal distribution is questionable for several reasons. First, many other distributions, such
as exponential, Poisson, Weibull-Rosin-Rammler, beta, or families of gamma or compound
gamma distributions \cite{Villermaux:2011ffa, Lefebre} capture to some extent the complexity of atomization physics.
Second, the geometrical standard deviation (GSD) of the log-normal fits to the many experimental measurements is relatively small (of the order of 1.2 \cite{Ling17} or 1.8 \cite{Marty:2015vs})  while the wide
range of scales in Fig. \ref{replot} seems to indicate a much larger GSD. Indeed
Nicas, Nazarroff \& Roberts \cite{Nicas:2017eo} obtain $\hat \sigma \simeq 8-9$. One explanation for the smaller
GSD in jet atomization studies, both numerical and experimental, is that the numerical or optical resolution is limited at the small scales. Indeed, as grid resolution is increased, the
observed GSD also increases \cite{Ling17}.
Third, many authors observe multimodal or bimodal distributions, that can be obtained for example by the superposition of several physical processes. This would arise in a very simple manner if the Taylor-Culick rim  route produced drops of a markedly different size than the holes-in-film route.

\begin{figure}[ht]
	\begin{center}
		%\vskip -100pt
		\includegraphics[width=0.6\linewidth] {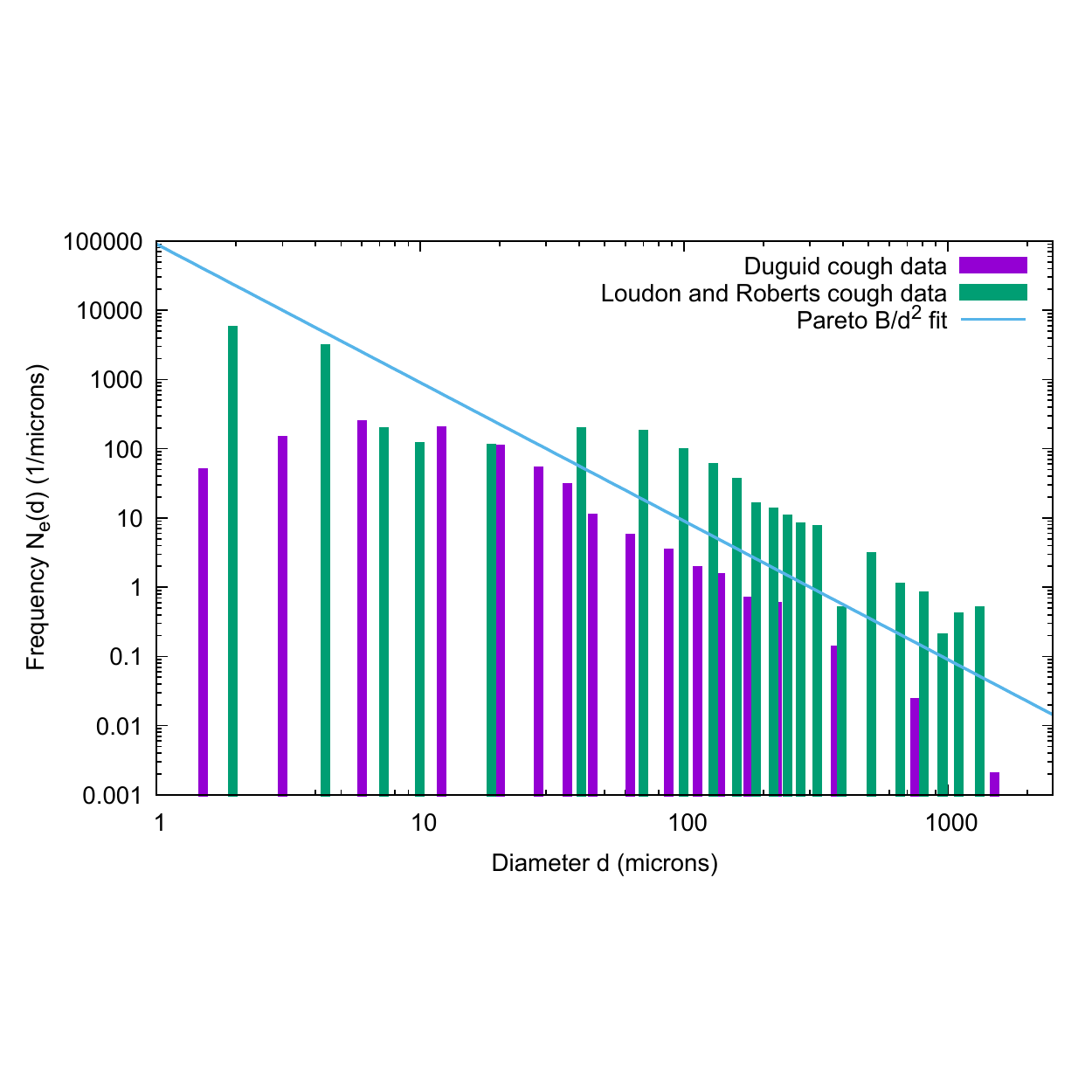}
	\end{center}
	%\vskip -60pt
	\caption{Frequency of droplet size distribution, replotted from Duguid \cite{Duguid:1946dw} and  Loudon \& Roberts \cite{loudon1967relation}. The Pareto distribution is also plotted. 
		\label{replot}}
\end{figure}

\begin{figure}
	\begin{center}
		%\vskip -40pt
		\includegraphics[width=0.6\linewidth] {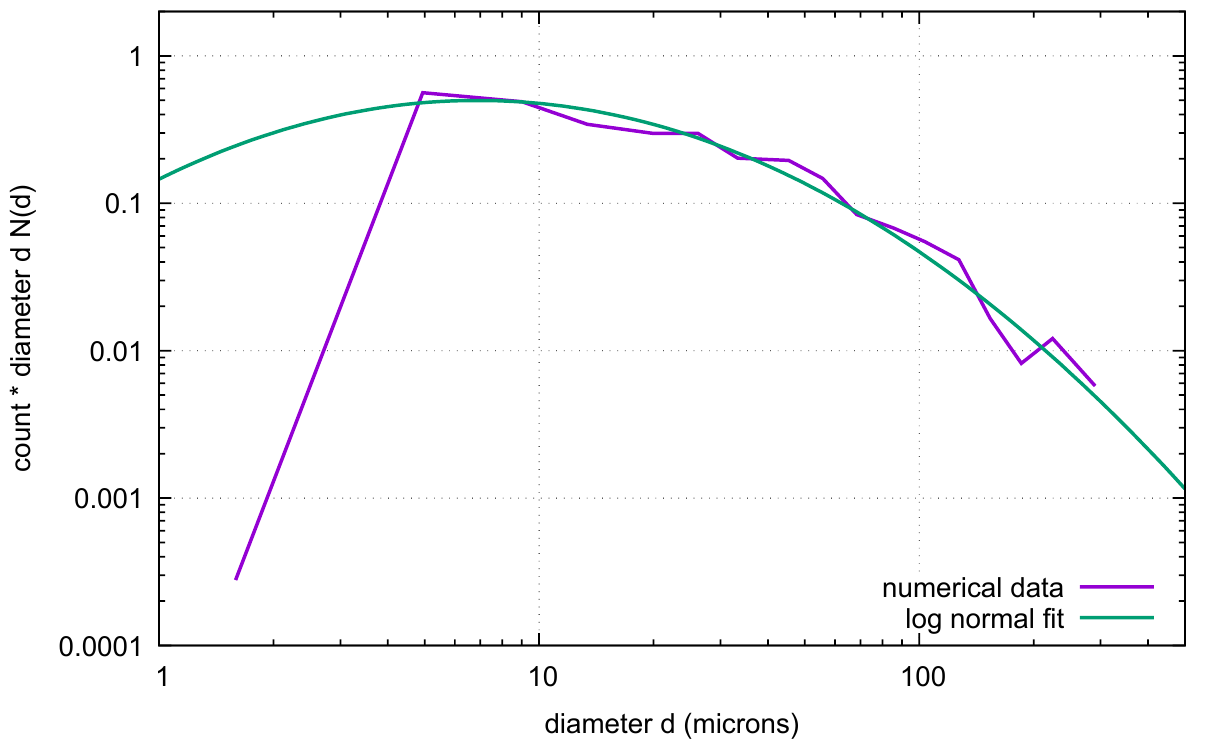}
		%\vskip -80pt
	\end{center}
	\caption{Count per droplet size, replotted from the numerical
		simulations of \cite{Ling17}.
		%in the same coordinates as Figure \ref{coughfit}.
		The $y$-axis is the count $N_i$ times the diameter
		$d_i$ in bin $i$. We took $d_i$ as the center of the bin.
		\label{ling2}}
\end{figure}

In order to elucidate this discrepancy, we take another look at the fit of the Duguid data
in Fig. \ref{coughfit}.
We replot the data that was provided in Table 3 of Duguid. Since the data are given as counts $N_i$ in bins defined by
the interval $(d_i, d_{i+1})$,
we approximate $N_e(d)$ at collocation points $d_{i+1/2}$ as $N_e(d_{i+1/2}) = N_i/(d_{i+1} - d_i)$,
with $d_{i+1/2} = (d_{i+1} + d_i)/2$.
We then plot $d \, N(d)$ in log-log coordinates in Fig. \ref{coughfit}, since
if plotted in the variables $x=\ln d$ and $y=\ln [d N_e(d)]$ the distribution (\ref{ln}) appears as a parabola. When one attempts to fit a parabola
between 2 and 50 $\mu m$, one obtains a log-normal distribution with $\hat \sigma = 0.7$ and
${\hat \mu} = \ln(12)$ (for diameters in microns).  However the data above 50 $\mu m$ are completely outside this distribution. If instead the whole range from 1 to 1000 $\mu m$ is fit to a log-normal distribution, one obtains a very wide log-normal or alternatively a Pareto distribution
of power 2
\be
\mathrm{Pareto \; Distribution:} \quad N_e(d) = \frac {B}{ d^2}
\ee
In Figs. \ref{replot} and
\ref{coughfit} we represent the Pareto distribution together with the Duguid  and Loudon \& Roberts data. It is especially clear from
Fig. \ref{replot} that if one does not trust either data
at $d < 50$ $\mu m$ then both data sets are well described by the Pareto distribution.

This however does not eliminate the possibility that more data with more statistical power could show deviations from Pareto, in particular as {\it multimodal} distributions. 
%On Fig. \ref{coughfit} we represent the statistical error bars near the maxima and minima, which are significant deviations.
Nevertheless, the multimodal deviation from the Pareto distribution is difficult to characterize and will not be pursued in what follows for the sake of simplicity.
\begin{figure}
	\centering
	\includegraphics[width=0.6\linewidth] {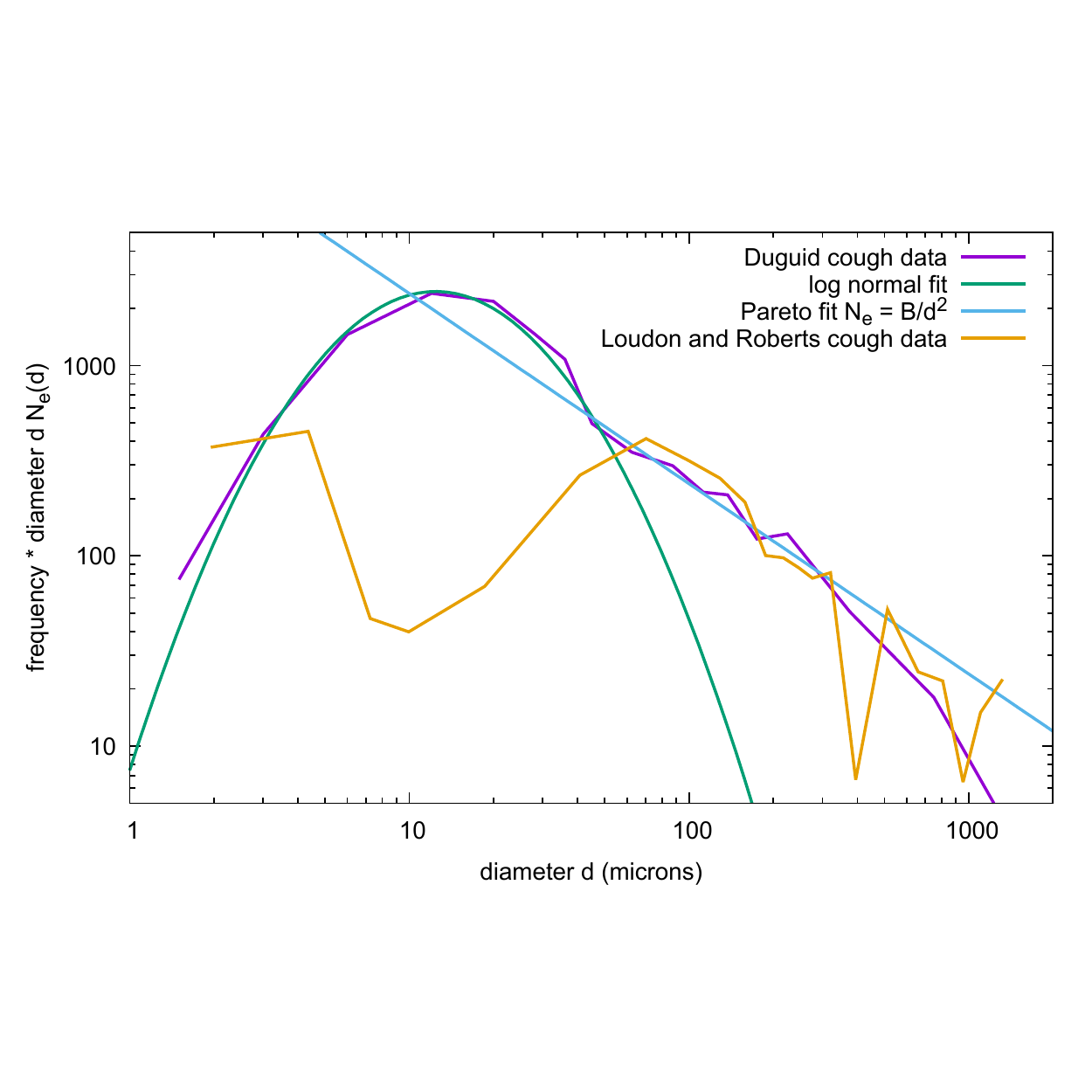}
	\caption{Fitting the data of Fig. \ref{replot} to log-normal and Pareto distributions,
		in the same coordinates as Fig. \ref{ling2}.
		The fit is adequate only up to 50 $\mu m$. As a result, only a fraction of the reliable
		data fits the log-normal.
		The Pareto distribution is a reasonable capture of the data in the 10 to 1000 $\mu m$ range.
		%The abcissa is the count $N_i$ times the diameter
		%$d_i$ in bin $i$. We took $d_i$ as the center of the bin in the original data.
		In the log-log coordinates the log-normal distribution appears as a parabola while the Pareto distribution is
		a straight line.
		%
		%The red vertical bars represent an estimate of the statistical error, equal to two standard deviations $\sigma (d_i N_i) = d_i \sqrt{N_i}$.
		\label{coughfit}
	}
\end{figure}
It is clear that the Pareto distribution cannot be valid at diameters that are either too large or too small. The equivalent diameter of the total mass of liquid being atomized is an obvious upper bound, but it is also very unlikely that droplets with $d > h$ where $h$ is the initial film thickness will be observed. It is reasonable to put this film thickness on the scale of $1$ $mm$ which corresponds to the upper bound on diameters in the data of Figs. \ref{replot} and \ref{coughfit}.
%We note however that the data at large diameter are 
The lower bound on droplet diameter is much harder to determine. {Exhalations are highly transient, or unsteady, processes involving complex multiscale geometry} \cite{scharfman2016visualization}, and thread breakup is a fractal multiscale process with satellite droplets \cite{Eggers97, tjahjadi1992satellite}. Going down in scale,
the fractal process repeats itself as long as continuum mechanics remains valid,
to around 1 $nm$. This would not be relevant for viral disease propagation as a lot of the relevant viruses have sizes {ranging from $O(10-100 \, nm)$, with an estimated size for SARS-CoV-2 ranging from 60-120 $nm$, for example}. If the smallest length scale is the thickness at which the thin liquid sheets will break, then
experimental observations in water \cite{opfer14} suggest a scale of  $O(100) \, nm$.
Other fluids, {including biological fluids or biologically contaminated fluids} such as those investigated in
\cite{poulain2018ageing, Wang:2018PRL, poulain2018bio}
may yield different length scales. 
%
%\Lnote{Here - for the following sentence -  it would be beneficial to be a bit careful, as viral load is not distributed equally in respiratory tract, so it is important not to imply that smaller size drops may lead automatically to lower loads, as we could have a preferential creation site for small drops, with localized higher concentration of virus when compared to larger ones - depending on type of respiratory infection. Maybe a sentence to nuance?} 
%For practical purposes one may also have to disregard any droplets that are so small that they contain an insufficient viral load. Thus one should emphasize that the lower bound of the distribution should be carefully selected in any practical situation. 

Based on the above considerations, we take a histogram of droplet sizes that reads
\be
N_e(d) = \begin{cases}
	\dfrac B {d^2}  \hbox{\hskip 10 pt \rm for \hskip 10 pt}  d_1 < d < d_2 \\
	0 \hbox{\hskip 10 pt \rm otherwise.}
\end{cases} \label{paretodist}
\ee
where $d_1$ is set to $O(100 \, nm)$  and $d_2$ to $O(1 \, mm)$ for simplicity.
The total volume of the droplets is 
\be
Q_{de} %= r C \int_{d_1}^{d_2} N_e(x) x^3 dx
= \dfrac {\pi \, B} {12} ( d_2^2-d_1^2) \approx \dfrac {\pi \, B} {12}  d_2^2 \, .
\ee
%where $r=\rho_l \pi /6$.
Since $d_1$ is four orders of magnitude smaller than $d_2$,
the total number of droplets is well approximated by
\be
{\cal N}_e =  \frac {12 Q_{de}} {\pi} \frac{1}{ d_1 \, d_2^2}
\ee
and the cumulative number of droplets $f(x) = N_e(d_1 \le d\le x)$, i.e., the number of droplets with diameter smaller than $x$, is very well approximated by
\be
f(x) = \int_{d_1}^x N_e(d) \, {\rm d} d \simeq {\cal N}_e \left( 1  - \frac{d_1}{x}\right).
\ee
so that $f(10 d_1)/{\cal N}_e = 90 \%$ of the droplets are of size less that 10 $d_1 \simeq 1 \, \mu m$. 
In other words, a {\it {numerical}} majority of the droplets are
near the lower diameter bound. On the other hand, a majority of the {\it{volume}} of
fluid is in the larger droplet diameters.

\subsection{Droplet velocities and ejections angles} \label{sec3.2}
The distribution of velocities and ejection angles has been investigated in the atomization experiments of \cite{descamps2008gas} which follow approximately the geometry of a high speed stream peeling by a gas layer. These experiments were qualitatively reproduced in the numerical simulations of \cite{Ling:2018hb}. To cite ref. \cite{descamps2008gas}, {``most of the ejection angles are in the range 0${}^\circ$ to 40${}^\circ$, however it occurs occasionally that the drops are ejected with angles as high as 60${}^\circ$''}.

On the other hand, there are to our knowledge no experimental data on the velocity of droplets, as they are formed in an atomizing jet, that could be used directly to estimate the ejection speed of droplets in exhalation. There are however numerical studies \cite{hoepffner11,jerome2013vortices} in the limit of very large Reynolds and Weber number. The group velocity of waves formed on a liquid layer below a gas stream has been estimated
by Dimotakis \cite{dimotakis86} as
\be
V_{de} \sim \left( \frac {\rho_p}{ \rho_d} \right)^{1/2} v_{pe} \, ,
\ee
where $\rho_d$ is droplet density. In \cite{hoepffner11,jerome2013vortices} it was shown that this was also the vertical velocity of the interface perturbation.
It is thus likely that this velocity plays a role at the end of the first instability stage of atomization. After this stage, droplets are detached and immersed in a gas stream of initial ejection velocity $v_{pe}$. Since the density ratio $\rho_p/\rho_d$ is $O(10^{-3})$, we expect the initial velocity of the ejected droplets at the point of their formation to be small.

As we show below, it is interesting to note that the large Reynolds number limit may apply at the initial injection stage to a wide range of droplets in the spectrum of sizes found above.
%%, which can be applied 
%%The speed with which droplets relax to the surrouding gas velocity
%%may be estimated as follows.
Indeed the ejection Reynolds number of a droplet ejected at a velocity $V_{de}$ in a surrounding air flow of velocity $v_{pe}$ is
\be
\Re_e =  \frac{|V_{de} - v_{pe}| d}{\nu_a} \, ,
\ee
where $\nu_a$ is the kinematic viscosity of the ejected puff of air (here taken to be the same as that of the ambient air). The largest Reynolds number is obtained for the upper bound of $d=1 \, mm$. For example, if the droplet's initial velocity is set to $V_{de} \approx 0$, and the air flow velocity in some experiments \cite{jama} is as high as 30 m/s, we can estimate the largest ejection Reynolds number to be $\Re_e \approx 2000$
and the Reynolds number will stay above unity for droplets down to micron size. But as the puff of air and the droplets move forward, the droplet Reynolds number rapidly decreases for the following reasons: (i) as will be seen in section~\ref{sec4.1} the puff velocity decreases  due to entrainment and drag, (ii) as will be seen in section~\ref{sec4.2.1} the droplet diameter {will decrease rapidly due to evaporation}, (iii) as will be seen in section~\ref{sec4.2.2} the time scale $\tau_V$ on which the droplet accelerates to the surrounding fluid velocity of the puff is quite small, and (iv) very large droplets quickly fall out of the puff and do not form part of airborne droplets. Thus, it can be established that droplets smaller than 100 $\mu m$ quickly equilibrate with the puff within the first few $cm$ after exhalation.

\section{Transport Stage} \label{sec4}
This section will consider the evolution of the puff of hot moist air with the droplets after their initial ejection. First in section~\ref{sec4.1} {we will present a simple modified model for the evolution of the puff of exhaled air, evaluating the effects of drag and the inertia of the droplets within it. This will enable us, in section~\ref{sec4.2} to discuss the evolution of the droplet size spectrum, velocity and temperature distributions, with simple first order models}. Additionally, section~\ref{sec4.3} will discuss the effect of non-volatiles on the droplet evolution and the formation of a fully evaporated droplet residue or aerosol particle.  Late-time turbulent dispersion of the virus-laden droplet residues, when the puff of air within which they are contained stops being a coherent entity, is then addressed in section~\ref{sec4.4}.    

\subsection{Puff Model} \label{sec4.1}
For the puff model we follow the approach of Bourouiba {\it {et al.}} \cite{Bourouiba2014}, but include the added effects of drag and the mass of the injected droplets. In addition, a perturbation approach is pursued to obtain a simple solution with all the added effects included. Fig. \ref{fig2} shows the evolution of the puff along with quantities that define the puff \cite{Bourouiba2014}. We define $t$ to be the time elapsed from exhalation and $s(t)$ to be distance traveled by the puff since exhalation. For analytical considerations we define the virtual origin to be at a distance $s_{e}$ from the real source in the backward direction and $t_e$ to be the time it takes for the puff to travel from the virtual origin to the real source. We define $t' = t+t_e$ to be time from the virtual origin and $s' = s+ s_e$ to the distance traveled from the virtual origin - their introduction simplifies the analysis.

From the theory of jets, plumes, puffs and thermals \cite{morton} the volume of the puff  exhaled  grows by entrainment. Bourouiba {\it {et al.}} \cite{Bourouiba2014} defined the puff to be spheroidal in shape with the transverse dimension to evolve in a self-similar manner as $r'(t') = \alpha s'(t')$, where $\alpha$ is related to entrainment coefficient. The volume of the puff is then $Q_p(t')= \eta r'^3(t') = \eta \alpha^3 s'^3(t')$ and the projected, or cross-sectional,  area of the puff $a(t')= \beta r'^2(t') = \beta \alpha^2 s'^2(t')$, where the constants $\eta$ and $\beta$ depend on the shape of the spheroid. For a spherical puff $\eta = 4 \pi/3$ and $\beta = \pi$.

As defined earlier, the ejected puff at the real source (i.e., at $t'=t_e$) is characterized by the volume $Q_{pe} = \eta \alpha^3 s_e^3$, momentum $M_{pe} = \rho_{pe}Q_{pe} v_{pe}$, buoyancy $B_{pe}=Q_{pe} (\rho_{a} - \rho_{pe})g$ and ejection angle $\theta_{e}$. From the assumption of self-similar growth we obtain the virtual origin to be defined as
\be
s_e = \left( \dfrac{Q_{pe}}{\eta} \right)^{1/3} \dfrac{1}{\alpha} \quad \mathrm{and} \quad t_e = \dfrac{\rho_{pe} \, Q_{pe}^{4/3}}{(4+C)\alpha M_{pe} \eta^{1/3}} \, ,
\ee
where the constant $C$ depends on the drag coefficient of the puff and will be defined below. If we assume a spherical puff with an entrainment factor $\alpha = 0.1$ \cite{morton}, the distance $s_e$ depends only on the ejected volume. Experimental measurements suggest $Q_{pe}$ to vary over the range 0.00025 to 0.0025 $m^3$. Accordingly, $s_e$ can vary from 0.39 to 0.84 $m$. Similar estimates of $t_e$ can be obtained for a spherical puff: as $Q_{pe}$ varies from 0.00025 to 0.0025 $m^3$ and as the ejected velocity varies from 1 to 10 $m/s$ the value of $t_e$ varies over the range 0.01 to 0.21 $s$.

The horizontal and vertical momentum balances in dimensional terms are
\be
\dfrac{\rd (M_d + M_p) \cos \theta}{\rd t'} = -\dfrac{1}{2} \rho_a \, a \, C_D \left(\dfrac{\rd s'}{\rd t'} \right)^2 \cos\theta \, ,
\ee 
\be
\dfrac{\rd (M_d + M_p) \sin \theta}{\rd t'} = B_{pe} -\dfrac{1}{2} \rho_a\, a \, C_D \left(\dfrac{\rd s'}{\rd t'} \right)^2 \sin\theta \, .
\ee
In the above $C_D$ is the drag coefficient of the puff and $M_d$ is the momentum of droplets within the puff. While the puff velocity decreases rapidly over time, the velocity of the larger droplets will change slowly. Note that in the analysis to follow we take the velocity of those droplets that remain within the puff to be the same as the puff velocity.

\begin{figure}[hbt!]		   	
	\begin{center}
		\includegraphics[width=0.6\linewidth]{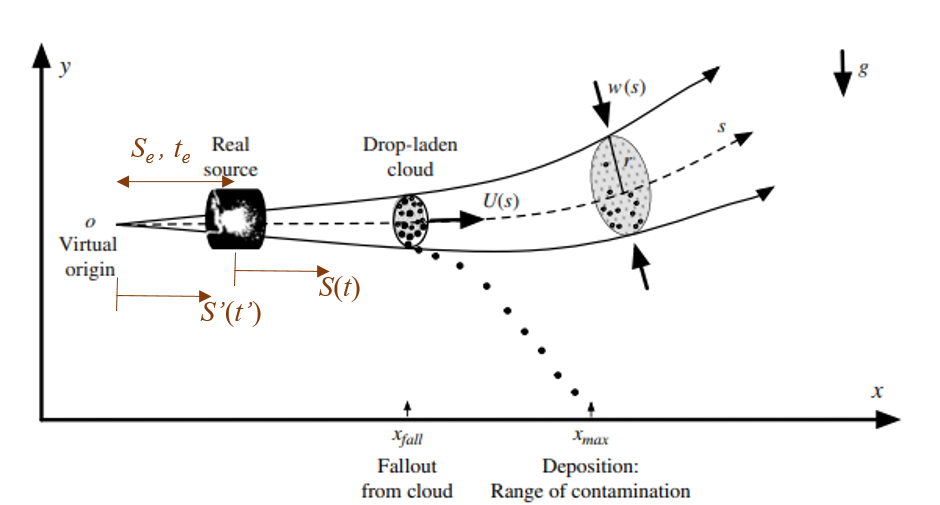}
	\end{center}   
	\caption{Evolution of a {typical cloud of respiratory multiphase turbulent droplet-laden air} following breathing, talking, coughing and sneezing activities. Image adapted from \cite{Bourouiba2014}.}
	\label{fig2}
\end{figure}

We use $s_e$ and $t_e$ as the length and time scales to define nondimensional quantities: $\tilde{s} = s'/s_e$ and $\tilde{t} = t'/t_e$. With this definition the virtual origin becomes $\tilde{t} =0$ and $\tilde{s} = 0$ and the real source becomes $\tilde{t} = 1$ and $\tilde{s} = 1$. In terms of non-dimensional quantities the governing momentum equations can be rewritten as
\be
\dfrac{\rd}{\rd \tilde{t}} \left[ \left( r_m \dfrac{\rd \tilde{s}}{\rd \tilde{t}} + \dfrac{1}{4} \dfrac{\rd \tilde{s}^4}{\rd\tilde{t}} \right) \cos \theta \right] = -C \, \tilde{s}^2 \left( \dfrac{\rd\tilde{s}}{\rd \tilde{t}} \right)^2 \cos \theta \, ,
\ee
\be
\dfrac{\rd}{\rd\tilde{t}} \left[ \left( r_m \dfrac{\rd \tilde{s}}{\rd \tilde{t}} + \dfrac{1}{4} \dfrac{\rd \tilde{s}^4}{\rd\tilde{t}} \right) \sin \theta \right] = A -C \, \tilde{s}^2 \left( \dfrac{\rd\tilde{s}}{\rd \tilde{t}} \right)^2 \sin \theta \, .
\ee
There are three nondimensional parameters: mass ratio of the initial ejected droplets to the initial air puff: $r_m = \rho_d Q_{de} /(\rho_p Q_{pe})$; the scaled drag coefficient: $C = C_D \beta /(2 \eta \alpha)$; and the buoyancy parameter: $A = B_{pe} t_e^2/(\rho_{pe} Q_{pe} s_e)$. In the above equations, $r_m$ is defined in terms of the mass of the initial ejected droplets. This is an approximation since some of the droplets exit the puff over time. Even though the droplet mass decreases due to evaporation, the associated momentum is not lost from the system since it remains within the puff. In any case, soon it will be shown that the value of $r_m$ is small and the role of ejected droplets on the momentum balance is negligible. It should also be noted that under Boussinesq approximation the small difference in density between the puff and that the ambient is important only in the buoyancy term. For all other purposes the two will be taken to be the same and as a result the time variation of puff density is not of importance (i.e., $\rho_p = \rho_{pe} = \rho_a$). 

The importance of inertia of the ejected droplets, drag on the puff and buoyancy effects can now be evaluated in terms of the magnitude of the nondimensional parameters. Typical experimental measurements of breathing, talking, coughing and sneezing indicate that the value of $r_m$ is smaller than 0.1 and often much smaller. Furthermore, as droplets  fall out continuously \cite{Bourouiba2014} from the turbulent puff, this ratio changes over time. Here we will obtain an upper bound on the inertial effect of injected droplets by taking the value of $r_m$ to be 0.1. 

The drag coefficient of a spherical puff of air is also typically small - again as an upper bound we take $C_D = 0.1$, which yields $C=0.375$ for a spherical puff. The value of the buoyancy parameter $A$ depends on the density difference between the ejected puff of air and the ambient, which in turn depends on the temperature difference. For the entire range of ejected volumes and velocities, the value of $A$ comes to be smaller than 0.01, for temperature differences of the order of ten to twenty degrees between the exhaled puff and the ambient.

Since all three parameters $r_m$, $C$ and $A$ can be considered as small perturbations, the governing equations can be readily solved in their absence to obtain the following classical expressions for the nondimensional puff location and puff velocity:
\be
\mathrm{when} \, (r_m=C=A = 0): \quad \tilde{s}(\tilde{t}) = \tilde{t}^{1/4} \quad \mathrm{and} \quad \tilde{v}(\tilde{t}) = \dfrac{d \tilde{s}}{d \tilde{t}} = \dfrac{1}{4} \tilde{t}^{-3/4} \, .
\ee
With the inclusion of the drag term the governing equations become nonlinear. Nevertheless, they allow a simple exact solution which can be expressed as
\be \label{eq_Conly}
\mathrm{when} \, (r_m=A = 0): \quad \tilde{s}(\tilde{t}) = \tilde{t}^{1/(4+C)} \, .
\ee 
Thus, as to be expected, the forward propagation of the puff slows down with increasing nondimensional drag parameter $C$. For small values of $C$ the above can be expanded in Taylor series as
\be
\mathrm{when} \, (r_m=A = 0): \quad \tilde{s}^4(\tilde{t}) = \tilde{t} - \dfrac{C}{4} \tilde{t} \ln(\tilde{t}) + \dfrac{C^2}{32} \tilde{t} (2 \ln(\tilde{t}) + \ln(\tilde{t})^2) + O(C^3) \, .
\ee
A comparison of the exact solution with the above asymptotic expansion shows its adequacy for small values of $C$. 

For small non-zero values of $r_m$, $C$ and $A$, the governing equations can be solved using regular perturbation. The result can be expressed as
\be
\tilde{s}^4(\tilde{t}) = \tilde{t} - \dfrac{C}{4} \tilde{t} \ln(\tilde{t}) + \dfrac{C^2}{32} \tilde{t} (2 \ln(\tilde{t}) + \ln(\tilde{t})^2) + 4 r_m (1 - \tilde{t}^{1/4}) - \dfrac{A \sin\theta_{e}}{2} (1 - \tilde{t})^2  \, 
\ee
and the above expression is accurate to $ O(C^3, r_m^2, A^2)$. Although the effect of buoyancy is to curve the trajectory of the puff, the leading order effect of buoyancy is to only alter the speed of rectilinear motion. Also, as expected, the effect of non-zero $r_m$ is to add to the total inertia and thereby slow down the motion of the puff. On the other hand, the effect of buoyancy is to slow down if the initial ejection is angled down (i.e., if $\theta_{e} < 0$) and to speed up if the ejection is angled up, provided the ejected puff is warmer than the ambient. 

The time evolution of the puff as predicted by the above analytical expression is shown in Fig. \ref{fig3}. Note that the point of ejection is given by $\tilde{t} = 1$, $\tilde{s} = 1$, and the initial non-dimensional velocity $\tilde{v}(\tilde{t}=1) = 1/4$. The results for four different combinations of $C$ and $r_m$ are shown. The buoyancy parameter has very little effect on the results and therefore is not shown. It should be noted that at late stages when the puff velocity slows down the effect of buoyancy can start to play a role as indicated in experiments and simulations. It can be seen that the effect of inertia of the ejected droplets, even with the upper bound of holding their mass constant at the initial value, has negligible effect. Only the drag on the puff has a significant effect in reducing the distance traveled by the puff. It can then be taken that the puff evolution to good accuracy can be represented by \eqref{eq_Conly}. Over a time span of 10 nondimensional units the puff has traveled about $0.7s_e$ and the velocity has dropped to about 15\% of the initial velocity.  By 100 nondimensional units the puff has traveled about $1.75 s_e$ and the velocity has dropped to about 2.5\% of the initial velocity.

\begin{figure}[hbt!]		   	
	\begin{center}
		\includegraphics[width=0.95\linewidth]{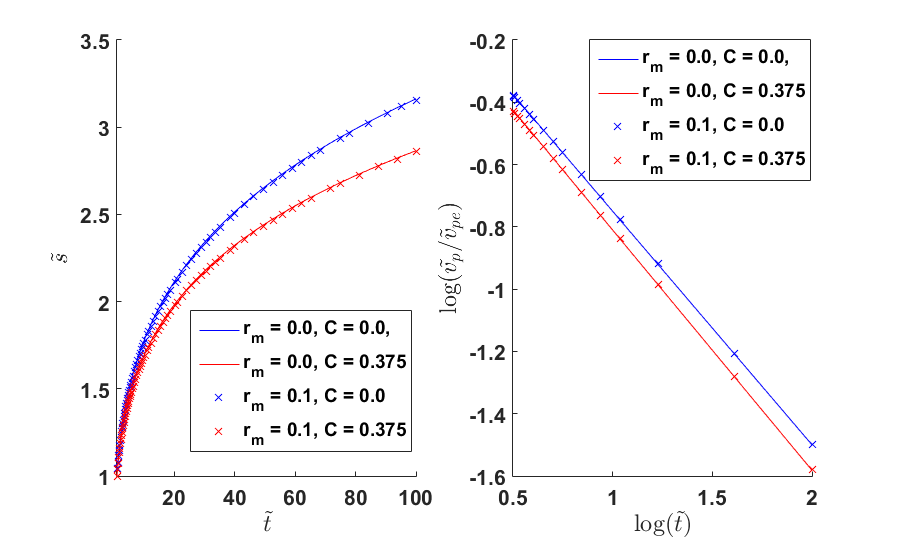}
	\end{center}   
	\caption{(a) Time evolution of a nondimensional puff distance $\tilde{s}$ for different combinations of $C$ and $r_m$. Note that at the source of ejection $\tilde{s}(\tilde{t}=1)=1$. (b) Time evolution of puff velocity $\tilde{v}_p$ scale by the initial velocity $\tilde{v}_{pe} = 1/4$ plotted on log scale. Thus, $y-axis$ reaching a value of -1 corresponding to a puff velocity of ten times smaller than the initial velocity at ejection.}
	%\Lnote{May be of interest to also show in log-log to highlight easily by eye scaling of 1/4 discussed in text? Kai's corrections}} 
	\label{fig3}
\end{figure}

\subsection{Droplet Evolution} \label{sec4.2}

The ejected droplets are made of a complex fluid that is essentially a mixture of oral fluids, including secretions from both the major and minor salivary glands. In addition, it is added up with several constituents of non-salivary origin, such as gingival crevicular fluid, exhalted bronchial and nasal secretions, serum and blood derivatives from oral wounds, bacteria and bacterial products, viruses and fungi, desquamated epithelial cells, other cellular components, and food debris \cite{Kaufman2002}. Therefore, it is not easy to determine precisely transport properties of the droplet fluid. Although surface tension is measured similar to that of water, viscosity can be one or two orders of magnitude larger \cite{Gershkovitch2002} making drops less coalescence prone \cite{roccon, soligo}.  In the present context, viscosity and surface tension might be of importance, because they can influence droplet size distribution specifically by controlling coalescence and breakage. These processes are important only during the ejection stage, and once droplets are in the range below $50 \, \mu m$, coalescence and break up processes are impeded, allowing us to model drops as single, non-interacting small spheres.

The ejected swarm of droplets is characterized by its initial size spectrum as given in \eqref{paretodist}. The time evolution of the spectrum of droplets that remain within the puff in terms of droplet size, velocity and temperature is the object of interest in this section. This evolution of the ejected droplets depend on the following four important parameters: the time scale $\tau_V$ on which the droplet velocity relaxes to the puff fluid velocity (in the absence of other forcing), the time scale $\tau_T$ on which the droplet temperature relaxes to the puff fluid temperature, the settling velocity $W$ of the droplet within the puff fluid, and the Reynolds number $\Re$ based on settling velocity. These quantities are given by \cite{bala-scaling1, ling-scaling2, ling-scaling3}
\be
\tau_V = \dfrac{\rho d^2}{18 \nu_p \Phi} \, , \quad \tau_T = \dfrac{\rho d^2 C_r}{6 \kappa_p Nu} \, , \quad  W = \tau_V g \, \quad \mathrm{and} \quad \Re = \dfrac{Wd}{\nu_p} \, ,
\ee
where $\rho \approx 1000 $ is the droplet-to-air density ratio, $C_r \approx 4.16$ is the droplet-to-air specific heat ratio, $g$ is acceleration due to gravity, $\nu_p$ and $\kappa_p$ are the kinematic viscosity and thermal diffusivity of the puff.  In the above, $\Phi = 1+ 0.15 \Re^{0.687}$ and $Nu = 2+ 0.6 \Re^{1/2} Pr^{1/3}$ are the finite Reynolds number drag and heat transfer correction factors; both of which simplify in the Stokes regime for drops smaller than about $50 \, \mu m$. Here we take the Prandtl number of air to be $Pr=0.72$. In the Stokes limit, the velocity and thermal time scales, and the settling velocity of the droplet increase as $d^2$, while Reynolds number scales as $d^3$. The value of these four parameters for varying droplet sizes is presented in Fig.~\ref{fig4}, where it is clear that the effect of finite $\Re$ becomes important only for droplets larger than 50 $\mu m$. For smaller droplets $\tau_V, \tau_T \ll 1 (s)$, $W \ll 1 (m/s)$, and $\Re \ll 1$. 

\begin{figure}[hbt!]		   	
	\begin{center}
		\includegraphics[width=0.6\linewidth]{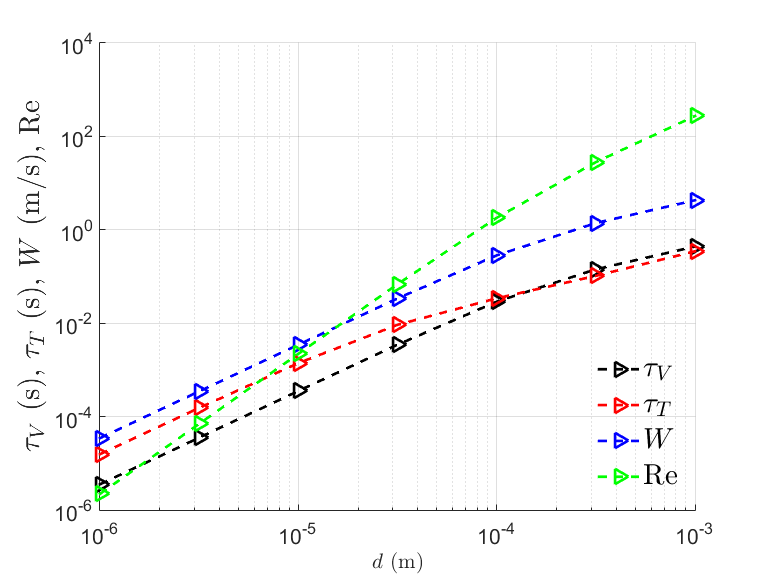}
	\end{center}   
	\caption{Dependence of velocity time scale $\tau_V$, thermal time scale $\tau_T$, still fluid settling velocity $W$ and Reynolds number of a settling droplet $\Re$ as a function of droplet diameter.}
	\label{fig4}
\end{figure}
 
\subsubsection{Droplet Size Controlled by Evaporation} \label{sec4.2.1}
%{\color{blue} Sometimes it is used $\tau_V$ others $\tau_p$ }
The size of the droplets under investigation is sufficiently small, and the swarm is dilute to prevent their coalescence, and the only way in which droplets change their size is via evaporation. According to the analysis of Langmuir \cite{Langmuir1918}, the rate of mass loss due to evaporation of a small sphere depends on the diffusion of the vapor layer away from the sphere surface, and under reasonable hypotheses \cite{Langmuir1918, Evans1945, Sazhin2006, Pirhadi}, it can be expressed as : 
\begin{equation}  
\label{dmdt1}
-\dfrac{\rd m}{\rd t} = \pi d D \rho_p Nu \ln(1+B_m),
\end{equation}
where, $m$ is the mass of a droplet of diameter $d$, $D$ is the diffusion coefficient of the vapor, $\rho_p$ is the density of puff air and $B_m = (Y_d - Y_p)/(1-Y_s)$ is the Spalding mass number, where $Y_d$ is mass fraction of water vapor at the droplet surface and $Y_p$ is mass fraction of water vapor in the surrounding puff. Under the assumption that $Nu$ and $B_m$ are nearly constant for small droplets, the above equation can be integrated to obtain the following law (mapping) for the evolution of the droplet:
\begin{equation}  \label{dmdt3}
d(t) = \mathcal{D} (d_{e},t) = \sqrt{d_e^2 - k' t},
\end{equation}
where $d_e$ is the initial droplet diameter at ejection and $k' = 4 D Nu \ln(1+B_m)/\rho$ has units of $m^2/s$ and thus represent an effective evaporative diffusivity. 
%the behavior of the drop radius evolution with time is shown in Figure~\ref{drop-time}. It is thus clear that the droplets evaporation rate is faster than the puff evolution rate (i.e. distance traveled) making the influence of droplet inertia on puff evolution more and more negligible. In addition, considering that the settling rate of the larger drops in the puff is order of less than a $cm/s$, we can also safely neglect the settling rate of drops. 
%
%\begin{figure}[h]
%	\centering
%	\includegraphics{empty.png}
%	\caption{\label{drop-time} Evolution of drop diameter with time}
%\end{figure}
%
It is important to observe that \eqref{dmdt1} would predict a loss of mass per unit area tending to infinity as the diameter of the drop tends to zero. This implies that the droplet diameter goes to zero in a finite time and we establish the result
\be \label{tevap}
d_{e,evap} \sim \sqrt{k'\, t} \,,
\ee 
which for any time $t$ yields a critical value of droplet diameter, and all droplets that were smaller, or equal,  at exhalation (i.e., $d_e \le d_{e,evap}$) would have fully evaporated by $t$. The only parameter is $k'$. Assuming $Nu =2$ and $D = 2.8 \times 10^{-5} m^2/s$, even for very small values of $B_m$, we obtain the evaporation time for a 10 $\mu m$ droplet to be less than a second. However, it appears that smaller than a certain critical size, the loss of mass due to evaporation slows down \cite{Evans1945}. This could partly be due to the presence of non-volatiles and other particulate matter within the droplet, whose effects were ignored in the above analysis, and will be addressed in section~\ref{sec4.3}. It seems that \eqref{dmdt1} can give reliable predictions for droplet diameter down to few $\mu m$ with much slower evaporation rates for smaller size. Irrespective of whether water completely evaporates leaving only the {non-volatile droplet residue}, or the droplet evaporation slows down, the important consequence on the evolution of the droplet size distribution is that it is narrower and potentially centered around micron size.

\subsubsection{Droplet Motion} \label{sec4.2.2}

We now consider the motion of the ejected droplets, while they rapidly evaporate. The equation of motion of the droplet is the Newton's law
\be
m \dfrac{\rd \bV_d}{\rd t} = -g (m - m_p) \bbe_z - 3\pi \rho_p \nu_p d \, \Phi \, (\bV_d - \bv_p) \, ,
\ee
where $\bbe_z$ is the unit vector along the vertical direction, $m_p$ is the mass of puff displaced by the droplet, $\bV_d$ and $\bv_p$ are the vector velocity of the droplet and the surrounding puff. Provided the droplet time scale $\tau_V$ is smaller than the time scale of surrounding flow, which is the case for droplets of diameter smaller than 50 $\mu m$, the above ODE can be perturbatively solved to obtain the following leading order solution \cite{bala-eea, ferry-implicit, BalachandarARF}
\be
\bV_d(t) = \bv_p(t) - W \bbe_z - \tau_V \dfrac{\rd \bv_p}{\rd t} \, .
\ee 
According to the above equation, the equilibrium Eulerian velocity of the droplet depends on the local fluid velocity plus the still fluid settling velocity $W$ of the droplet plus the third term that arises due to the inertia of the droplet. Though at ejection the droplet speed is smaller than the surrounding gas velocity, as argued in section~\ref{sec3.2}, the droplets quickly accelerate to approach the puff velocity. In fact, since the puff is decelerating (i.e. $|\rd \bv_p /\rd t| < 0$), the droplet velocity will soon be larger than the local fluid velocity. As long as the droplet stays within the puff, the velocity and acceleration of the surrounding fluid can be approximated by those of the puff as $|\bv_p| = \rd s/\rd t$ and $|\rd \bv_p /\rd t| = \rd^2 s/\rd t^2$. This allows evaluation of the relative importance of the third term (inertial slip velocity) in terms of the puff motion, which is given in \eqref{eq_Conly} as \cite{ling-scaling2}
\be
\dfrac{\tau_V |\rd \bv_p /\rd t|}{|\bv_p|} = \left( \dfrac{3+C}{4+C}\right) \dfrac{\tau_V}{t_e} \dfrac{1}{\tilde{t}} \, .
\ee
This ratio takes its largest value at the initial time of injection and then decays as $1/\tilde{t}$. Using the range of possible values of $t_e$ given earlier, this ratio is small for a wide range of initial droplet sizes. {We thus confirm} that for the most part droplet inertia can be ignored in its motion, and the droplet velocity can be taken to be simply the sum of local fluid velocity and the still fluid settling velocity of the droplet.

\subsubsection{Droplet Exit From the Puff}\label{sec4.2.3}
While the effect of buoyancy on the puff was shown to be small, the same cannot be said of the droplets. The vertical motion of a droplet with respect to the surrounding puff, due to its higher density, is dependent only on the fall velocity $W$, which scales as $d^2$, which in turn decreases as given in \eqref{dmdt3} due to evaporation. The droplet's gravitational settling velocity can be integrated over time to obtain the distance over which it falls as a function of time. We now set this fall distance (left hand side) equal to the puff radius (right hand side) to obtain
\be \label{texit}
\dfrac{\rho g}{18 \nu_a} \left( d_{e,exit}^2 \, t - \dfrac{1}{2} k' t^2 \right) = \alpha s_e \left( \dfrac{t+t_e}{t_e}\right)^{1/(4+C)} \, ,
\ee
where we have set the droplet diameter at exhalation to be $d_{e,exit}$, indicating the fact that a droplet of initial diameter equal to $d_{e,exit}$ has fallen by a distance equal to the puff size at time $t$. Thus all larger droplets of size $d_e > d_{e,exit}$ have fallen out of the puff by $t$ and we have been referring to these as the {{exited droplets}}.

The two critical initial droplet diameters, $d_{e,evap}$ and $d_{e,exit}$ are plotted in Fig.~\ref{fig5}a as a function of $t$. The only other key parameter of importance is $k'$ whose value is varied from $10^{-12}$ to $10^{-6} \, m^2/s$. In evaluating $d_{e,exit}$ using \eqref{texit}, apart from the property values of water and air, we have used the nominal values of $\alpha = 0.1$, $s_e = 0.5 \, m$ and $t_e = 0.05 \, s$ (as an example). The solid lines correspond to $d_{e,exit}$, which decreases with increasing $t$ and for each value of $k'$ there exits a minimum $d_e$ below which there is no solution to \eqref{texit} since the droplet fully evaporates before falling out of the puff. The dotted lines correspond to $d_{e,evap}$, which increases with $t$. The intersection of the two curves is marked by the solid square, which corresponds to the limiting time $t_{lim}(k')$, beyond which the puff contains only fully-evaporated droplet residues containing the viruses. Correspondingly we can define a limiting droplet diameter $d_{e,lim}(k')$. Given sufficient time, all initially ejected larger droplets (i.e., $d_e > d_{e,lim}$) would have fallen out of the puff and all smaller droplets (i.e., $d_e \le d_{e,lim}$) would have evaporated to become droplet residues. 

\begin{figure}[hbt!]		   	
	\begin{center}
		\includegraphics[width=0.49\linewidth]{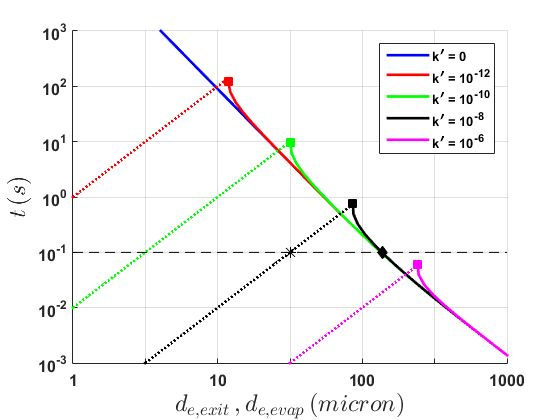}
		\includegraphics[width=0.49\linewidth]{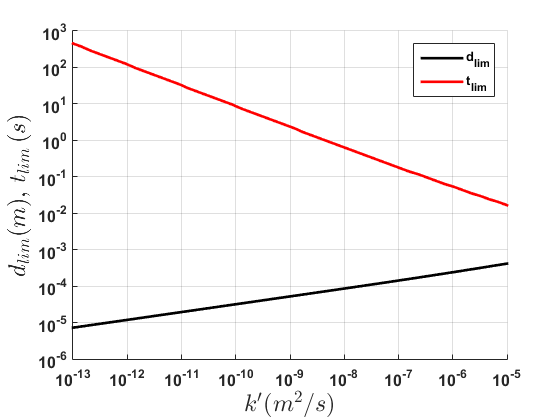}
	\end{center}   
	\caption{(a) Variation of $d_{e,evap}$ as a function of $t$ plotted as dotted lines and variation of $d_{e,exit}$ as a function of $t$ plotted as solid lines. The results for different values of $k'$ are shown in different colors. Note for $k' \rightarrow 0$ it takes infinite time for droplet evaporation. The solid square symbols denote the limiting droplet size $d_{e,lim}$ below which evaporation dominates and above which settling effect dominates. The corresponding time $t_{lim}$ is important since beyond this time all the droplets of initial diameter greater than $d_{e,lim}$ have fallen out of the puff and all the droplets below this size that remain within the puff are fully-evaporated droplet residues. For any time $t < t_{lim}$, we can identify a $d_{e,evap}$ (marked by the * for $t=0.1 \, s$ and $k'= 10^{-8} \, m^2/s$) below which all droplets have become residues, and a $d_{e,exit}$ (marked by solid diamond) above which all droplets have fallen out of the puff. All ejected droplets of intermediate initial size (i.e., $d_{e,evap} < d_e < d_{e,exit}$) remain within the puff partially evaporated. (b) Variation of $d_{e,lim}$ and $t_{lim}$ as a function of $k'$.}
	%\Lnote{possibly add graph converting the k' values into RH values for a given indoor temperature T - for more general audience to understand implication for indoor or outdoor spaces? I could not find the part marked with a "*" for t = 3s, is that a glitch in image?}}
	\label{fig5}
\end{figure}

At times smaller than the limiting time (i.e., for $t < t_{lim}$) we have the interesting situation of some droplets falling out of the puff (exited droplets), some still remaining as partially evaporated airborne droplets, and some fully-evaporated to become droplet residues. This scenario is depicted in Fig.~\ref{fig5}a with an example of $t = 3 \, s$ for $k'= 10^{-10} \, m^2/s$ plotted as a dash line.
%\Lnote{I could not find example of 3s with dashed line in Fig 8a}. 

%From equations \eqref{tevap} and \eqref{texit} we can now define the two limiting diameters
%\be
%d_{evap} = \sqrt{k' t} \quad \mathrm{and} \quad d_{exit} = \left[ \dfrac{18 \nu_a}{\rho g} \dfrac{\alpha s_e}{t} \left(\dfrac{t+t_e}{t_e}\right)^{1/(4+c)} + \dfrac{1}{2} k' t \right]^{1/2} \, .
%\ee
%By time $t$, when less than $t_{exit}$, initially ejected droplets of size larger than $d_{exit}$ would have fallen out of the puff, droplets smaller than $d_{evap}$ would have evaporated to an aerosol, and droplets of intermediate size would still remain in the puff as partially evaporated state, whose current size can easily be obtained from \eqref{dmdt3}. The variation of $t_{lim}$ and $d_{e,lim}$ as a function of $k'$ is presented in Fig.~\ref{fig5}b. It is clear that as $k'$ varies over a wide range $t_{lim}$ ranges from 0.01 s to 450 s, and correspondingly $d_{e,lim}$ varies from 415 to 7 microns.

\subsection{Effect of Non-Volatiles} \label{sec4.3}
There can be significant presence of non-volatile material such as mucus, bacteria and bacterial products, viruses and fungi, and food debris in the ejected droplets \cite{Kaufman2002}. However, the fraction of ejected droplet volume $Q_{de}$ that is made up of these non-volatiles varies substantially from person to person. The presence of non-volatiles alters the analysis of the previous sections in two significant ways. First, each ejected droplet, as it evaporates, will reach a final size that is dictated by the amount of non-volatiles that were initially in it. The larger the  droplet size at initial ejection, the larger will be its final size after evaporation, since it contains a larger amount of non-volatiles. If $\psi$ is the volume fraction of non-volatiles in the initial droplet, the final diameter of the droplet residue after complete evaporation of volatile matter (i.e., water) will be 
\be
d_{dr} = d_e \psi^{1/3} \, .
\ee
{This size depends on the initial droplet size and  composition. Note that even a small, for example 1\%, non-volatile composition results in $d_{dr}$ being around 20\% of the initial ejected droplet size. It has also been noted that the evaporation of water can be partial, depending on local conditions in the cloud or environment. We simply assume the fraction $\psi$ to also account for any residual water retained within the droplet residue. }

The second important effect of non-volatile is to reduce the rate of evaporation. As evaporation occurs at the droplet surface, a fraction of the surface will be occupied by the non-volatiles reducing the rate of evaporation. For small values of $\psi$, the effect of non-volatiles is quite small only at the beginning. The effect of non-volatiles will increase over time, since the volume fraction of non-volatiles increases as the volatile matter evaporates. Because of this ever decreasing evaporation rate, it may take longer for a droplet to decrease from its ejection diameter of $d_e$ to its final droplet residue diameter of $d_{dr}$, than what is predicted by \eqref{dmdt3}. It should be noted that intermittency of turbulence and heterogeneity of vapor concentration and droplet distribution within the puff will influence the evaporation rate \cite{Ernst, Meunier2017, Eaton1994}}.   Nevertheless, for simplicity, {and for the purposes of the present first order mathematical framework, we use} the $d^2$-law given in \eqref{dmdt3}, but with a smaller value of effective $k'$ to account for the effect of non-volatiles and turbulence intermittency. This approximation is likely to be quite accurate in describing the early evolution of the droplet. Only at late stages as the droplet approaches its final diameter $d_{dr}$, the $d^2$-law will be in significant error.

Applying the analysis of the previous sections, taking into account the presence of non-volatiles, we separate the two different time regimes of $t \le t_{lim}$ and $t\ge t_{lim}$. In the case when $t \le t_{lim}$, we have three types of droplets: (i) exited droplets whose initial size at injection is greater than $d_{e,exit}$, (ii) droplets of size at ejection smaller than $d_{e,evap}$ that have completely evaporated to become droplet residues of size $d_{dr}$ and (iii) intermediate size airborne droplets that are within the puff and still undergoing evaporation.
We assume an equation of the form \eqref{texit} to approximately apply even in the presence of non-volatiles. With this balance between fall distance of a droplet and the puff radius we obtain the following expression
\be
d_{e,exit} = \left[ \dfrac{18 \nu_a}{\rho g} \dfrac{\alpha s_e}{t} \left(\dfrac{t+t_e}{t_e}\right)^{1/(4+C)} + \dfrac{1}{2} k' t \right]^{1/2} \, .
\ee
The corresponding limiting diameter of complete evaporation can be obtained from setting $d = d_{e,evap} \psi^{1/3}$ and $d_e = d_{e,evap}$ in \eqref{dmdt3} as
\be
d_{e,evap} = \sqrt{\dfrac{k' t}{1-\psi^{2/3}}} \, .
\ee
While the above two estimates are in terms of the droplet diameter at injection, their current diameter at $t$ can be expressed as
\be
d_{evap} = d_{e,evap} \psi^{1/3} \quad \mathrm{and} \quad d_{exit}^2 = d^2_{e,exit} - k' t \, .
\ee

Form the above expressions, we define $t_{lim}$ to be the time when $d_{e,exit} = d_{e,evap}$, which in terms of current droplet diameter becomes $d_{exit} = d_{evap}$. Beyond this limiting time (i.e., for $t > t_{lim}$) the droplets can be separated into only two types: (i) exited droplets whose initial size at injection greater than $d_{e,exit} = d_{e,evap}$, and (ii) droplets of size at ejection smaller that have become droplet residues. The variation of $t_{lim}$ and $d_{e,lim}$ as a function of $k'$ is presented in Fig.~\ref{fig5}b. It is clear that as $k'$ varies over a wide range, $t_{lim}$ ranges from 0.01 s to 450 s, and correspondingly $d_{e,lim}$ varies from 415 to 7 $\mu m$.

\subsubsection{Droplet Size Spectrum Within the Puff}
We now put together all the above arguments to present a predictive model of the droplet concentration within the puff. The initial condition for the size distribution is set by the ejection process discussed in Section 3, and the simple Pareto distribution given in \eqref{paretodist} provides an accurate description. Based on the analysis of the previous sections, we separate the two different time regimes of $t \le t_{lim}$ and $t\ge t_{lim}$.

In the case when $t \le t_{lim}$ the droplet/aerosol concentration (or the number per unit volume of the puff) can be expressed as
\be \label{spec1}
\mbox{If} \;\; t \le t_{lim}: \quad \phi(d,t) = \begin{cases} 
	\dfrac{1}{Q(t)} \, N_e(d \psi^{-1/3})  &\mbox{for} \quad d \le d_{evap} \\
	\dfrac{1}{Q(t)} \, N_e\left( \sqrt{d^2 + k' t} \right) &\mbox{for} \quad d_{evap} \le d \le d_{exit} \\
    0 & \mbox{for} \quad d \ge d_{exit} \end{cases} \, ,
\ee
where we have recognized the fact that equation \eqref{dmdt3} is the mapping $\mathcal{D}$ between the current droplet size and its size at injection. Due to the turbulent nature of the puff, the distribution of airborne droplets and residues is taken to be uniform within the puff. Quantities such as $\tilde{s}$, $d_{evap}$ and $d_{exit}$ are as they have been defined above and the pre-factor $1/Q(t)$ accounts for the expansion of the puff volume. In the case of $t \ge t_{lim}$, the droplet number density spectrum becomes
\be \label{spec2}
\mbox{If} \;\; t \ge t_{lim}: \quad \phi(d,t) = \begin{cases} 
	\dfrac{1}{Q(t)} \, N_e(d \psi^{-1/3})  &\mbox{for} \quad d \le d_{lim} \\
		0 & \mbox{for} \quad d \ge d_{lim} \end{cases} \, ,
\ee
and only droplet residues remain within the puff. Here, the size of the largest droplet residue within the puff is related to its initial unevaporated droplet size as $d_{lim} = d_{e,lim} \psi^{1/3}$, and the plot of $d_{e,lim}$ as a function of $k'$ for a specific example case of puff and droplet ejection was shown in Fig.~\ref{fig5}b.

\subsubsection{Droplet Temperature}
In this subsection we will briefly consider droplet temperature, since it plays a role in determining saturation vapor pressure and the value of $k'$. Following Pirhadi \eal \cite{Pirhadi} we write the thermal equation of the droplet as
\be
m C_{pw} \dfrac{\rd T_d}{\rd t} = \pi k_p \, d \, Nu \dfrac{\ln (1+B_m)}{B_m} \left( T_p - T_d \right) + L \dfrac{\rd m}{\rd t} \, ,
\ee
where $C_{pw}$ is specific heat of water, $k_p$ is thermal conductivity of the puff air, $L$ is the latent heat of vaporization, $T_d$ and $T_p$ are the temperatures of the droplet and the surrounding puff. The first term on the right accounts for convective heat transfer from the surrounding air and the second term accounts for heat needed for phase change during evaporation.

It can be readily established that the major portion of heat required for droplet evaporation must come from the surrounding air through convective heat transfer. The equilibrium Eulerian approach \cite{bala-eea} can again be used to obtain the asymptotic solution of the above thermal equation and the droplet temperature can be explicitly written as
\be
T_d(t) \approx T_p(t) + \dfrac{L}{\pi k_p \, d \, Nu} \dfrac{B_m}{\ln(1+B_m)} \dfrac{\rd m}{\rd t} - \tau_T \dfrac{B_m}{\ln(1+B_m)} \dfrac{\rd T_p} {\rd t} \, ,
\ee
where $\tau_T$ is the thermal time scale of the droplet that was introduced earlier. The second term on the right is negative and thus contributes to the droplet temperature being lower than the surrounding puff. Simple calculation with typical values shows that the contribution of the third term is quite small and can be ignored. As a result, the temperature difference between the droplet and the surrounding is largely controlled by the evaporation rate $\rd m/\rd t$, which decreases over time. Again, using the properties of water and air, and typical values for $Nu$ and $B_m$, we can evaluate the temperature difference $T_p - T_d$ to be typically a few degrees. Thus, the evaporating droplets need to be only a few degrees cooler than the surrounding puff for evaporation to continue.

\subsection{Late-Time Turbulent Dispersion} \label{sec4.4}
When the puff equilibrates with the surrounding and its velocity falls below the ambient turbulent velocity fluctuation, the subsequent dynamics of the droplet cloud  is governed by turbulent dispersion. This late-time evolution of the droplet cloud depends on many factors that characterize the surrounding air. This is where the difference between a small enclosed environment such as an elevator or an aircraft cabin or an open field matters, along with factors such as cross breeze and ventilation. A universal analysis of late-time evolution of the droplet residue cloud is thus not possible, due to problem-specific details. The purpose of this brief discussion is to establish a simple scaling relation to guide when the puff evolution model presented in the above sections gives way to advection and dispersion by ambient turbulence. It should again be emphasized that the temperature difference between the puff fluid containing the droplet residue cloud and the ambient air may induce buoyancy effects, which for model simplicity will be taken into account as part of turbulent dispersion.

We adopt the classical scaling analysis of Richardson \cite{Richardson}, according to which the radius of a droplet cloud, in the inertial range, will increase as the $3/2$ power of time as given by
\be
r_{lt}^2(t) = c' \, \epsilon \, (t+t_0)^3 \, , \label{rich}
\ee 
where $c'$ is a constant, $\epsilon$ is the dissipation rate, which will be taken to be a constant property of ambient turbulence, and $t_0$ is the time shift required to match the cloud size at the transition time between the above simple late time model and the puff model. In the above, the subscript $lt$ stands for late-time behavior of the radius of the droplet-laden cloud. We now make a simple proposal that there exists a transition time $t_{tr}$, below which the rate of expansion of the puff as given by the puff model is larger than $dr_{lt}/dt$ computed from the above expression. During this early time, ambient dispersion effects can be ignored in favor of the puff model. But for $t > t_{tr}$ droplet-laden cloud's ambient dispersion becomes the dominant effect. 

The constants $t_0$ and $t_{tr}$ can be obtained by satisfying the two conditions: (i) the size of the droplet-laden cloud given by \eqref{rich} at $t_{tr}$ matches the puff radius at that time given by $\alpha s_e ((t_{tr}+t_e)/t_e)^{1/(4+C)}$, and (ii) the rate of expansion of the droplet-laden cloud by turbulent dispersion matches the rate of puff growth given by the puff model. This latter condition can be expressed as
\be
\dfrac{3}{2} \sqrt{c' \epsilon} (t_{tr}+t_0)^{1/2} = \dfrac{\alpha s_e}{(4+C)} \dfrac{1}{t_e^{1/(4+C)}} \left({t_{tr}+t_e}\right)^{-\frac{3+C}{4+C}} \, .
\ee
From these two simple conditions we obtain the final expression for the transition time as
\be
t_{tr} = \left( \dfrac{2 \alpha^{2/3} s_e^{2/3}}{3 (4+C) ({c' \epsilon})^{1/3}} \right)^{\frac{3(4+C)}{(10+3C)}} \dfrac{1}{t_e^{2/(10+3C)}} - t_e \, .
\ee
Given a puff, characterized by its initial ejection length and time scales $s_e$ and $t_e$, and the ambient level of turbulence characterized by $\epsilon$, the value of transition time can be estimated. If we take entrainment coefficient $\alpha = 0.1$, the constant $C = 0$, and typical values of $s_e = 0.5 \, m$ and $t_e = 0.05 \, s$, we can estimate $t_{tr} = 1.88 \, s$ for a dissipation rate of $c' \epsilon = 10^{-5} \, m^2/s^3$. The transition time $t_{tr}$ increases (or decreases) slowly with decreasing (or increasing) dissipation rate.   Thus, the early phase of droplet evaporation described by the puff model is valid for $O(1) \, s$, before being taken over by ambient turbulent dispersion. 

However, it must be stressed that the scaling relation of Richardson is likely an over-estimation of ambient dispersion, as there are experimental and computational evidences that suggest that the power-law exponent in \eqref{rich} is lower than $3$ \cite{Okubo}. But it must be remarked that even with corresponding changes to late-time turbulent dispersion, the impact on transition time can be estimated to be not very large. Also, it must be cautioned that according to classical turbulent dispersion theory, during this late-time dispersal, the concentration of virus-laden droplet residues within the cloud will not be uniform, but will tend to decay from the central region to the periphery. Nevertheless, for sake of simplicity here we assume \eqref{rich} to apply and we take the droplet residue distribution to be uniform.

According to above simple hypothesis, the effect of late-time turbulent dispersion on number density spectrum is primarily due to the expansion of the could, while the total number of droplet residues within the cloud remains the same. Thus, the expressions \eqref{spec1} and \eqref{spec2} still apply. However the expression for the volume of the cloud must be appropriately modified as
\be
\tilde{Q}(t) = \begin{cases}
	\eta \alpha^3 s_e^3 \left( \dfrac{t+t_e}{t_e} \right)^{3/(4+C)} & \mathrm{for} \quad t \le t_{tr} \\
	\eta (c' \epsilon)^{3/2} \, (t+t_0)^{9/2} & \mathrm{for} \quad t \ge t_{tr} \, .
\end{cases} \label{Qt}
\ee
The location of the center of the expanding cloud of droplets can be still taken to be given by the puff trajectory $s(t)$, which has considerably slowed down during late-time dispersal. The strength of the above model is in its theoretical foundation and analytical simplicity. But, the validity of the approximations and simplifications must be verified in applications to specific scenario being considered. {For example, considering variability in composition, turbulence intermittency, initial conditions of emissions and the state of the ambient, direct observations show that the transition between puff dominated and ambient flow dominated fate of respiratory droplets vary from O(1-100 \, s) \cite{jama}.}

\section{Inhalation Stage}
This section will mainly survey the existing literature on issues pertaining to what fraction of the droplets and aerosols at any location gets inhaled by the recipient host, and how this is modified by the use of masks. These effects modeled as inhalation and filtration efficiencies will then be incorporated into the puff-cloud model. The pulmonary ventilation (breathing) has a cyclic variation that varies markedly with age and metabolic activities.  The intensity of breathing (minute ventilation) is expressed in $L/min$ of inhaled and exhaled air.   For the rest condition, the ventilation rate is about 5-8 $L/min$ and increases to about 10-15 $L/min$ for mild activities.  During exercise, ventilation increases significantly depending on age and metabolic needs of the activity. 

In the majority of earlier studies on airflow and particle transport and deposition in human airways, the transient nature of breathing was ignored for simplification and to reduce the computational cost. Haubermann \eal \cite{Haubermann} performed experiments on a nasal cast and found that particle deposition for constant airflow is higher than those for cyclic breathing.  Shi \eal \cite{Shi} performed simulations on nanoparticle depositions in the nasal cavity under cyclic airflow and found that the effects of transient flow are important. Grgic \eal \cite{Grgic} and Horschler \eal \cite{Horschler} performed experimental and numerical studies, respectively, on flow and particle deposition in a human mouth-throat model, and the human nasal cavity.  Particle deposition in a nasal cavity under cyclic breathing condition was investigated by Bahmanzadeh \eal \cite{Bahmanzadeh}, Naseri \eal \cite{Naseri1}, and Kiasadegh \eal \cite{Kiasadegh}, where the unsteady Lagrangian particle tracking was used.  They found there are differences in the predicted local deposition for unsteady and equivalent steady flow simulations. In many of these studies, a sinusoidal variation for the volume of air inhaled is used.  That is
\be
Q_{in} = Q_{max} \sin (2 \pi t/T) \, .
\ee
Here $Q_{max}$ is the maximum flow rate, and $T=4 \, s$ is the period of breathing cycle for an adult during rest or mild activity.   The period of breathing also changes with age and the level of activity. Haghnegahdar \eal \cite{Haghnegahdar} investigated the transport, deposition, and the immune system response of the low-strain Influenza A Virus IAV laden droplets.  They noted that the shape of the cyclic breathing is subject dependent and also changes with nose and mouth breathing.  They provided an eight-term Fourier series for a more accurate descriptions of the breathing cycle. The hygroscopic growth of droplets was also included in their study. 

Analysis of aspiration of particles through the human nose was studied by Ogden and Birkett \cite{Ogden} and Armbruster and Breuer \cite{Armbruster}. Accordingly, the aspiration efficiency $\eta_a$ is defined as the ratio of the concentration of inhaled particles to the ambient concentration.   Using the results of earlier studies and also his works, Vincent \cite{Vincent} proposed a correlation for evaluating the inhalability of particles.  That is, the aspiration efficiency $\eta_a$ of particles smaller than 100 $\mu m$ is given as, 
\be
\eta_a = 0.5 \left[ 1+\exp(-0.06d) \right]     \quad \mathrm{for} \quad    d < 100 \, \mu m
\ee
Here, $d$ is an aerodynamic diameter of particles.  While the above correlation provides the general trend that larger particles are more difficult to inhale, it has a number of limitations. It was developed for mouth-breathing with the head oriented towards the airflow direction with speeds in the range of $1 \, m/s$ to $4 \, m/s$.
%In addition, it does not account for the influence of thermal plume.   
The experimental investigation of aerosol inhalability was reported by Hsu and Swift \cite{Hsu}, Su and Vincent \cite{Su1, Su2}, Aitken \eal \cite{Aitken}, and Kennedy and Hinds \cite{Kennedy}.    Dai \eal \cite{Dai} performed in-vivo measurements of inhalability of large aerosol particles in calm air and fitted their data to several correlations.  For calm air condition, they suggested,
\be
\eta_a = \begin{cases}
4.57 + 1.06 (\log d)^2 - 4.40 \log(d) \quad   \mathrm{For \; rest \; condition} \, , \\
4.16 + 0.97 (\log d)^2 - 4.10 \log(d) \quad   \mathrm{For \; moderate \; exercise} \, .
\end{cases}
\ee

\begin{figure}		   	
	\begin{center}
		\includegraphics[width=0.7\linewidth]{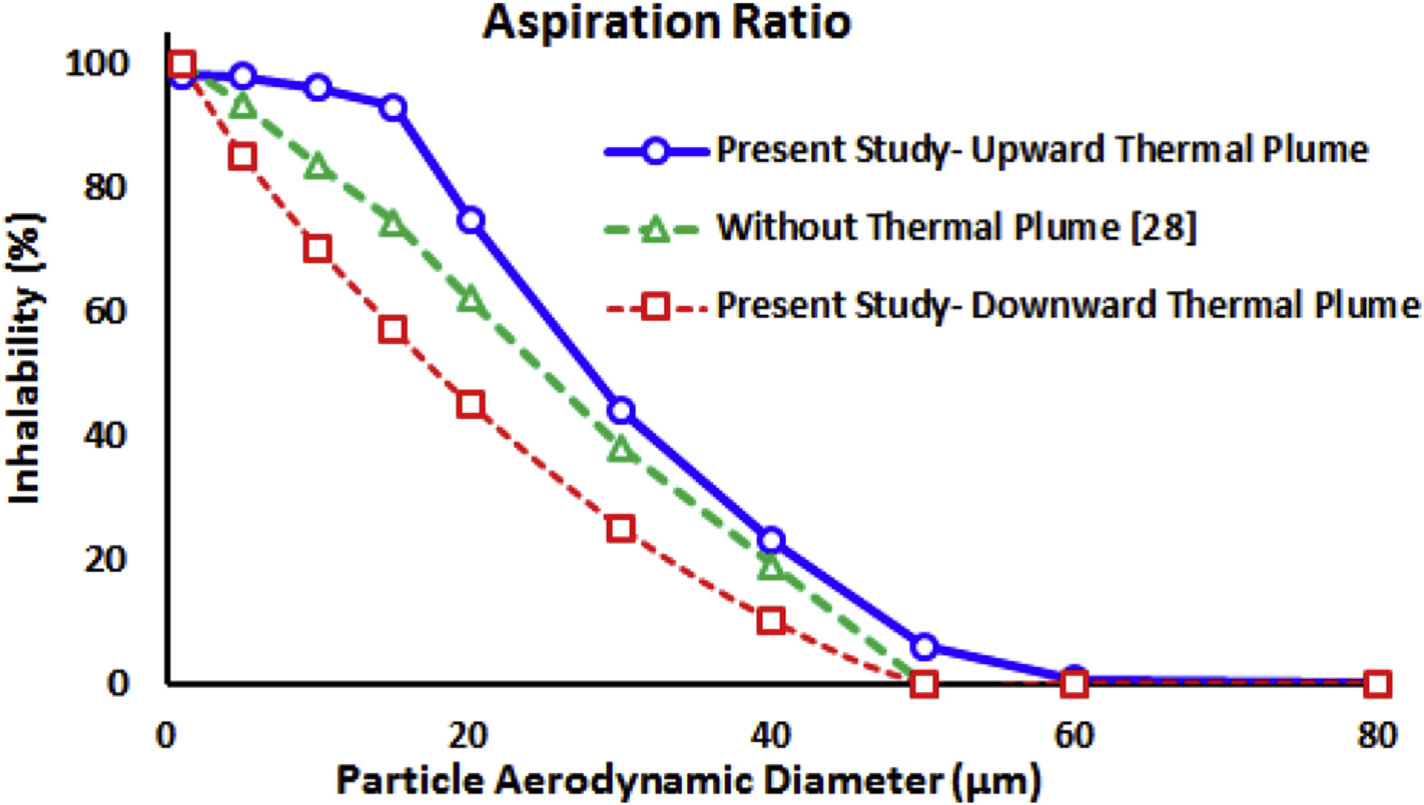}
	\end{center}   
	\caption{Influence of thermal plume on aspiration efficiency \cite{Naseri1}.}
	\label{Gfig1}
\end{figure}

Computational modeling of inhalability of aerosol particles were reported by many researchers \cite{King, Inthavong1, Inthavong2, Naseri1, Millage, Chen}.   Interpersonal exposure was studied by \cite{Gao, He}.    The influence of thermal plume was studied by Salmanzadeh \eal \cite{Salmanzadeh}.   Nasiri \eal \cite{Naseri2} performed a series of computational modeling and analyzed the influence of the thermal plume on particle aspiration efficiency when the body temperature is higher or lower than the ambient.  Their results are reproduced in Figure~\ref{Gfig1}.  Here the case that the body temperature $T_b =26.6 \, ^\circ C$ and the ambient temperature $T_a =21.3 \, ^\circ C$ (Upward Thermal Plume) and the case that $T_b = 32.2 \, ^\circ C$ and $T_a =40.0 \, ^\circ C$ (Downward Thermal Plume) are compared with the isothermal case studied by Dai \eal \cite{Dai}.   It is seen that when the body is warmer than the surrounding, the aspiration ratio increases.   When the ambient air is at a higher temperature than the body, the inhalability decreases compared to the isothermal case. In light of the results of the previous section it can be concluded that at a distance of $O(1) \, m$ the ejected droplets have sufficiently reduced in size that these $O(1) \, \mu m$ aerosols have near perfect inhalability. {However, recall that this estimation must be adjusted to account for the possible slow down of evaporation due to variability in the initial conditions of emissions and the ambient, intermittency of cloud turbulence and drop concentration. Thus, the above conclusion represents a lower bound of timescale and distance of transition to fully evaporated droplet residues from the cloud model \cite{jama}.}

%Their results are reproduced in Fig.~\ref{Gfig1}.

\begin{figure}		   	
	\begin{center}
		\includegraphics[width=0.5\linewidth]{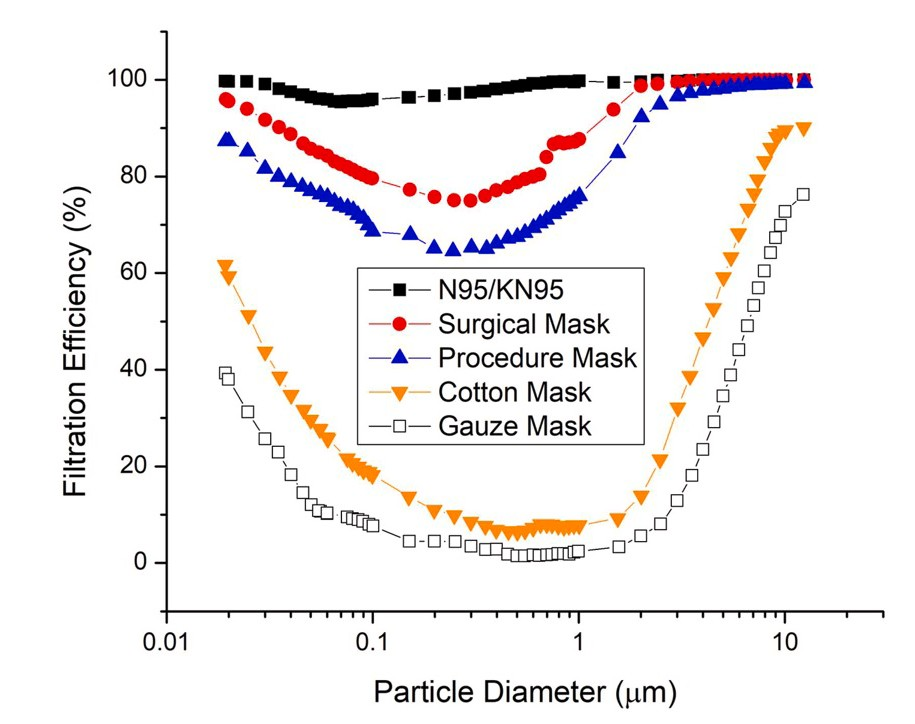}
	\end{center}   
	\caption{Filtration efficiency of different respiratory masks under normal breathing conditions \cite{Zhang, Feng}.}
	\label{Gfig2}
\end{figure}

\subsection{Respiratory Face Masks}
Using a respiratory face mask is a practical approach against exposure to airborne viruses and other pollutants.  Among the available facepiece respirators, N95, and surgical masks are considered to be highly effective \cite{Grinshpun, Loeb}. N95 mask has a filtration efficiency of more than 95\% in the absence of face leakage \cite{NIOSH, Qian}.   Surgical masks are used extensively in the hospital and operating rooms \cite{Lipp}.  Nevertheless, there have been concerns regarding their effective filtration of airborne bacteria and viruses \cite{Balazy, Lee, Loeb}.  There is often discomfort in wearing respiratory masks for extended durations that increases the risk of spread of infection. The breathing resistance of a mask is directly related to the pressure drop of the filtering material.   The efficiency of respiratory masks varies with several factors, including the intensity and frequency of breathing as well as the particle size \cite{Zou}.   The filtration efficiencies of different masks under normal breathing conditions, as reported by Zhang \eal \cite{Zhang} and Feng \eal \cite{Feng}, are shown in Figure~\ref{Gfig2}.  It is seen that the filtration efficiencies of different masks vary significantly, with N95 having the best performance, which is followed by the surgical mask.   It is also seen that all masks could capture large particles.  The N95, Surgical, and Procedure masks remove aerosols larger than a couple of microns.  Cotton and Gauze masks capture a major fraction of particles larger than 10 $\mu m$.   The capture efficiency of all masks also shows an increasing trend as particle size becomes smaller than 30 $nm$ due to the effect of the Brownian motion of nanoparticles.  Figure~\ref{Gfig2} also shows that the filtration efficiencies of all respiratory masks drop for the particle sizes in the range of 80 $nm$ to about 1 $\mu m$.  This is because, in this size range, both the inertia impaction and the Brownian diffusion effect are small, and the mask capture efficiency reduces. Based on these results, and the earlier finding that most ejected droplets within the cloud have become sub-micron-sized aerosol particles by about $O(1-10) \, m$ distance, it can be stated that only professional masks such as N95, Surgical, and Procedure masks provide reliable reduction in the inhaled particles. {Hence, the importance for healthcare workers to have access to high-grade respirators upon entering a room or space with infectious patients \cite{jama}}.

It should be emphasize that the concentration that a receiving host will inhale $(\phi_{inhaled})$ depends on the local concentration in the breathing zone adjusted by the aspiration efficiency given by Equations (40) and (41) (or plotted in Figure 9).   When the receiving host wears a mask, an additional important correction is needed by multiplying by a factor $(1-\eta_f)$, where $\eta_f$ is the filtration efficiency plotted in Figure 9.  That is, 
\be
\phi_{inhaled}(d,t) = \phi(d,t) \, \eta_a  (1 - \eta_f) \, ,
\ee
where $\phi(d,t)$ is the droplet residue concentration at the breathing zone given in \eqref{spec1} or \eqref{spec2}.
It is seen that the concentration of inhaled droplet larger than 10 microns  significantly decreases when the mask is used.  But the exposure to smaller droplets, particularly, in the size range of 100 $nm$ to 1 $\mu m$ varies with the kind of mask used.

\section{Discussion on Current Assumptions and Sample Analysis}
The object of this section is to put together the different models of the puff and droplet evolution described in the previous sections, underline their simplifications, and demonstrate their ability to make useful prediction. Such results under varying scenarios can then be potentially used for science-based policy making, such as establishing multi-layered social distancing guidelines and other safety measures. In particular, we aim at modeling the evolution of the puff and the concentration of airborne droplets and residues that remain within the cloud so that the probability of potential transmission can be estimated.  

As discussed in section~\ref{sec4.2}, the virus-laden droplets exhaled  by an infected host will undergo a number of transformations before reaching the next potential host. 
%Some of the droplets  settle down. {When not accounting for intermittency of turbulence and its effect on phase change, i.e. when using  the $d^2$-law, droplets of smaller settling speed that remain within the puff will undergo rapid evaporation with consequent size reduction}. Once droplets have  reached the fully evaporated {droplet residue, or aerosol state,} they will remain airborne for long times. 
To prevent transmission, current safety measures impose a safety distance of two meters.  Furthermore, cloth masks are widely used by public and their effectiveness has been shown to be questionable for droplets and aerosols of size about a micron. The adequacy of these common recommendations and practices can be evaluated by investigating the concentration of airborne droplets and residues at distances larger than one meter and the probability of them being around a micron in diameter, since such an outcome will substantially increase the chances of transmission. In the following we will examine two effects: the presence of small quantities of non-volatile matter in the ejected drops that remain as droplet residues after evaporation, and the adequacy of the log-normal or Pareto distribution to quantify the number of droplets in the lower diameter classes. 
%Finally, we will present a simple model which can be used to estimate the concentration of potentially contagious droplets in the puff and in confined environments.

\subsection{Current Predictions}
Let us consider the situation of speaking or coughing, whose initial puff volume and momentum are such that they yield $s_e \simeq 0.5 \, m$ and $t_e \simeq 0.05 \, s$. Under this specific condition, as shown in Figure~\ref{fig3} the puff travels about 1 $m$ in about 5 $s$ \footnote {The distance traveled can be upwards of 7-8 meters in a few seconds for sneezes, emitted with speeds on the order of 10-30 m/s \cite{jama}}. For this simple example scenario, we will examine our ability to predict airborne droplet and residue concentration, as an important step towards estimating potential for airborne transmission in situations commonly encountered.

%Considering that in the initial stage the puff will contain a large number of droplets, it is our aim here to evaluate how the size distribution of the ejected drops evolves. The size distribution of drops is a crucial figure which determine the airborne transmission potentials. 
%Xie, X., Li, Y., Chwang, A. T. Y., Ho, P. L., & Seto, W. H. (2007). How far droplets can move in indoor environmentsâ€“revisiting the Wells evaporationâ€“falling curve. Indoor air, 17(3), 211-225.
In most of the countries, current guidelines are based on the work by Xie \eal \cite{xie2007}, who revisited previous guidelines by \cite{Wells} with improved evaporation and settling models. They identified the possibility that, due to evaporation, the droplets quickly become aerosolized before reaching a significant distance and thus may represent a minor danger for transmission due to their minimal virus loading. 
\begin{figure}[h]
	\centering
	\includegraphics[ width=7cm]{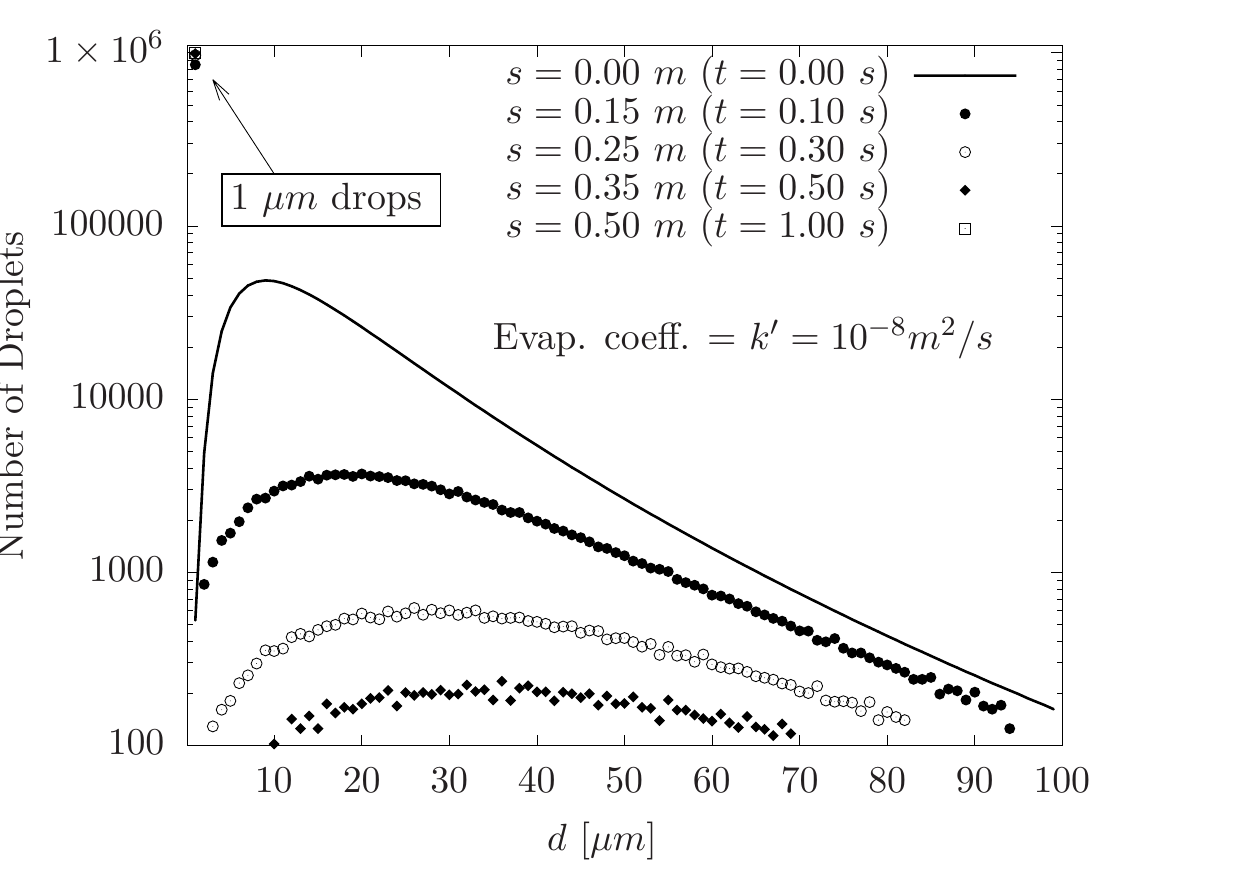}
	\caption{\label{spectra1} Evolution of the drop size distribution spectra according to the currently used evaporation models \cite{Wells, xie2007}.}
\end{figure}
This scenario is shown in Figure~\ref{spectra1}, where we present the evolution of the drop size spectrum while droplets are transported by the ejected puff. The initial droplet size distribution is taken to be that measured by Duguid \cite{Duguid:1946dw} modeled with a log-normal distribution, which in the Monte-Carlo approach is randomly sampled with one million droplets divided into one thousand diameter classes. Each droplet is then followed while evaporating and falling. The evaporation model is taken to be \eqref{dmdt3} with the effective diffusion coefficient estimated as $k' \simeq 1\cdot 10^{-8} \, m^2/s$. This value is computed under the assumption that drops are made of either pure water or a saline solution \cite{xie2007} and that air has about 98\% humidity. Therefore, this is an environment unfavorable to evaporation and consequently drop size reduction happens relatively slowly. However, from the figure it is clear that, even in this extreme case, after few tens of centimeters, and within a second, all droplets have evaporated down to a size  below $10 \, \mu m$. This is in line with the predictions of Xie \eal \cite{xie2007}. Naturally, if the air is dryer, the effective evaporation coefficient will be larger (even as large as $k' \simeq 10^{-5} \, m^2/s$) and the droplet size spectrum will evolve even faster, leaving virtually all droplets to be smaller than $1 \, \mu m$ in the puff. {Recall that intermittency of turbulence with the puff can create clusters of droplets and concentration of vapor and thereby significantly alter the evaporation rate \cite{Ernst, Meunier2017, Eaton1994}}.  Hence, our estimate of evaporation time is a lower bound, as governed by the $d^2$-law \eqref{dmdt3}.

\subsection{Influence of Non Volatile Matter}
\begin{figure}[h]
	\centering
	\includegraphics[ width=7cm]{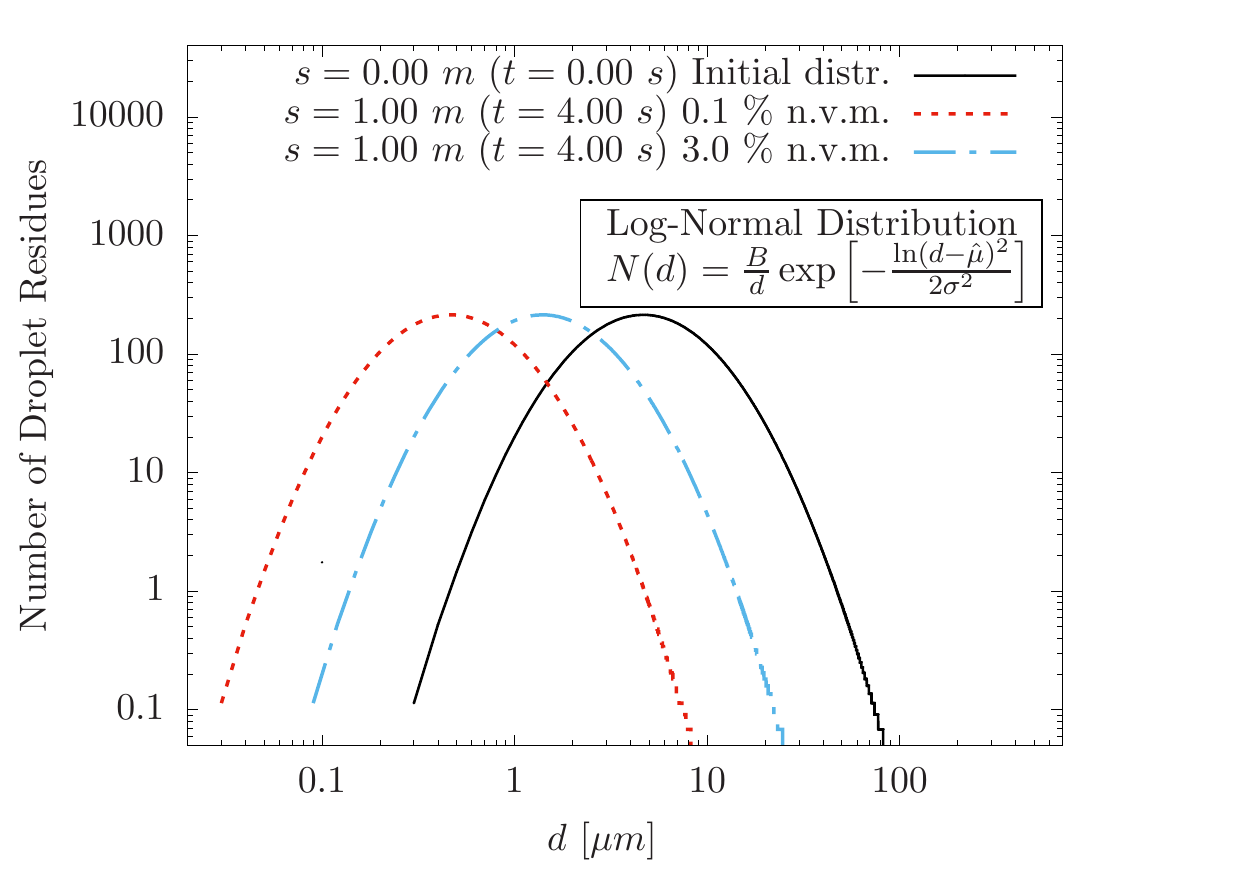} \includegraphics[ width=7cm]{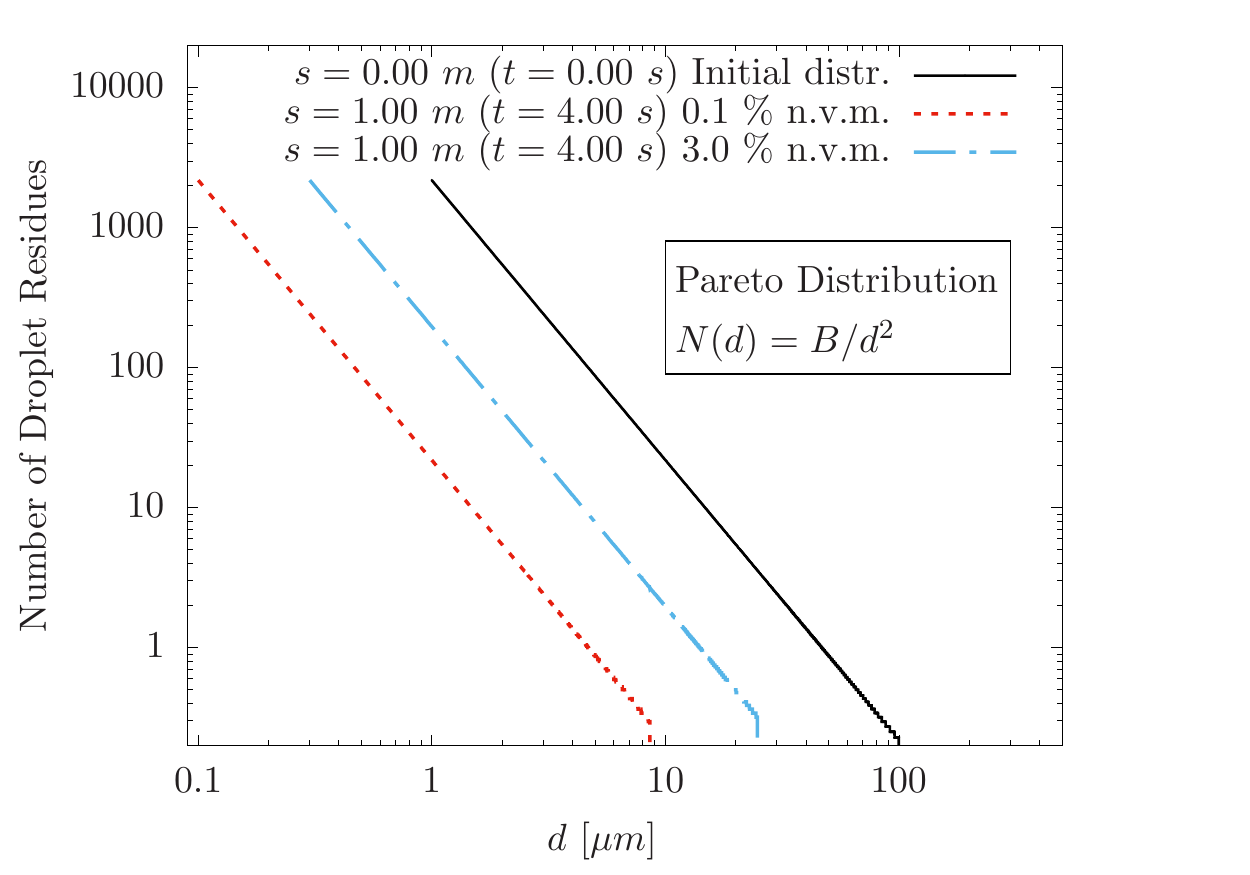}
	\caption{\label{spectra2}Influence of small quantities of non volatile matter on the final drop size distribution.}
\end{figure}

As discussed in section~\ref{sec4.3} there is current consensus that droplets ejected during sneezing or coughing contain, in addition to water, other biological and particulate non-volatile matter. Specifically, viruses themselves are just large protein chains of size almost $0.1 \, \mu m$. Here we will examine the evolution of droplet size distribution in the presence of non-volatile matter. It will be clear in the following, that in this case, even a small amount of non-volatile matter plays an important role with the evaporation coefficient being a minor factor in deciding how fast the final state is reached. In Figure~\ref{spectra2}, we show the final distribution of droplets under two scenarios where the initially ejected droplets contain $0.1 \%$ and $3.0 \%$ of non-volatile matter. In Figure~\ref{spectra2}a, the initial drop size distribution is modeled as a log-normal distribution (i.e. as in Fig.~\ref{spectra1})  whereas in Figure~\ref{spectra2}b, the initial drop size distribution is modeled according to the Pareto distribution with initial droplet size varying  between $1$ and $100 \, \mu m$. This range is smaller than that suggested earlier in section~\ref{sec3}. However drops that are larger than 100 $\mu m$ fall out of the cloud and therefore are not important for airborne transmission and droplets initially smaller than 1 $\mu m$ have much smaller viral load. Here ``final droplet size distribution'' indicates the number of droplets that remain within the puff after all the larger droplets have fallen out and all others have completed their evaporation to become droplet residues. This final number of droplet residues as a function of size does not vary with time or distance.

The size distribution is computed here as in Figure~\ref{spectra1}, with a random sampling from the initial log-normal or Pareto distribution. As before, these computations used an evaporation coefficient of $k' = 10^{-8} \, m^2/s$. However, there are two important differences: each droplet is allowed to fall vertically according to its time-dependent settling velocity, $W$, which decreases over time as the droplet evaporates. Integration of the fall velocity over time provides the distance traveled by the droplet relative to the puff. Droplets whose fall distance exceeds the size of the puff are removed from consideration. Second, each droplet that remains within the puff evaporates to its limiting droplet residue size that is dictated by the initial amount of non-volatile matter contained within the droplet. For $\psi = 0.1 \%$ non-volatile matter, the final aerosol size cannot decrease below $10 \%$ of the initial droplet diameter, whereas for $3.0 \%$ of non-volatile matter, the final droplet size cannot decrease below $30 \%$ of the initial diameter. From Fig.~\ref{spectra2}, it is clear that when eveporation is complete the drop size distribution rigidly shifts towards smaller diameters, with a cut-off upper diameter due to the settling of large drops. Essentially, it is clear that the initial number of viruses that were in droplets of size smaller than $d_{e,exit}$ still remain within the cloud almost unchanged, representing a more dangerous source of transmission than predicted by the conventional assumption of near-full evaporation.  Again, it is important to note that the final droplet size distribution is established rapidly even with the somewhat lower effective evaporation diffusivity of  $k' = 10^{-8} \, m^2/s$, {and when not accounting for the effect of localized moisture of the cloud in further reducing the rate}.  Figure \ref{spectra2} also illustrates the important difference in the drop size distribution. The Pareto distribution will predict a much larger number of drops in the micron and sub-micron range, possibly the most dangerous for both inhalation efficiency and filtration inefficiency.

\subsection{Sample Model Estimation of Airborne transmission}
\begin{figure}[h]
	\centering
	\includegraphics[ width=7cm]{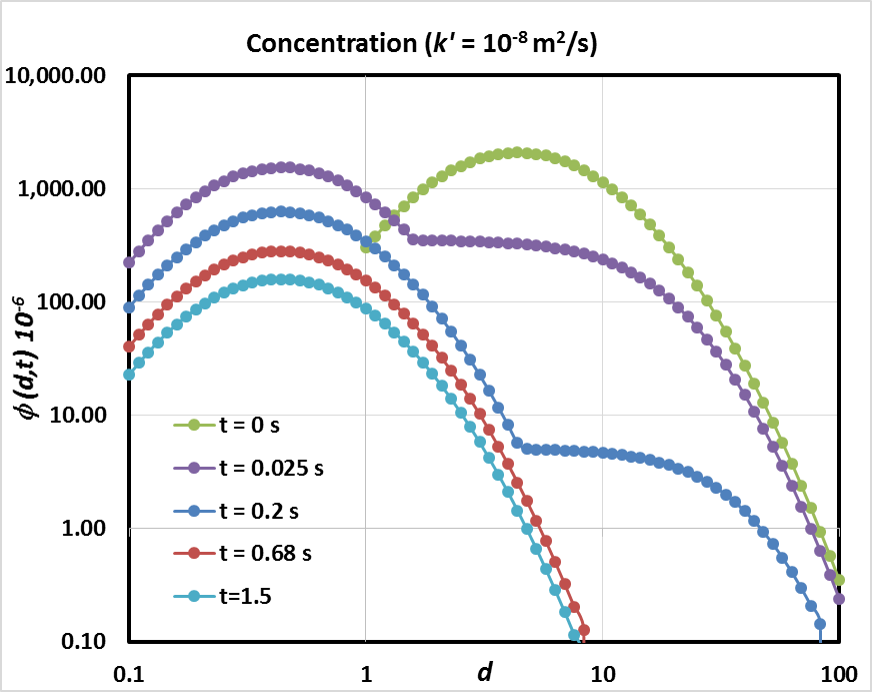} \includegraphics[ width=7cm]{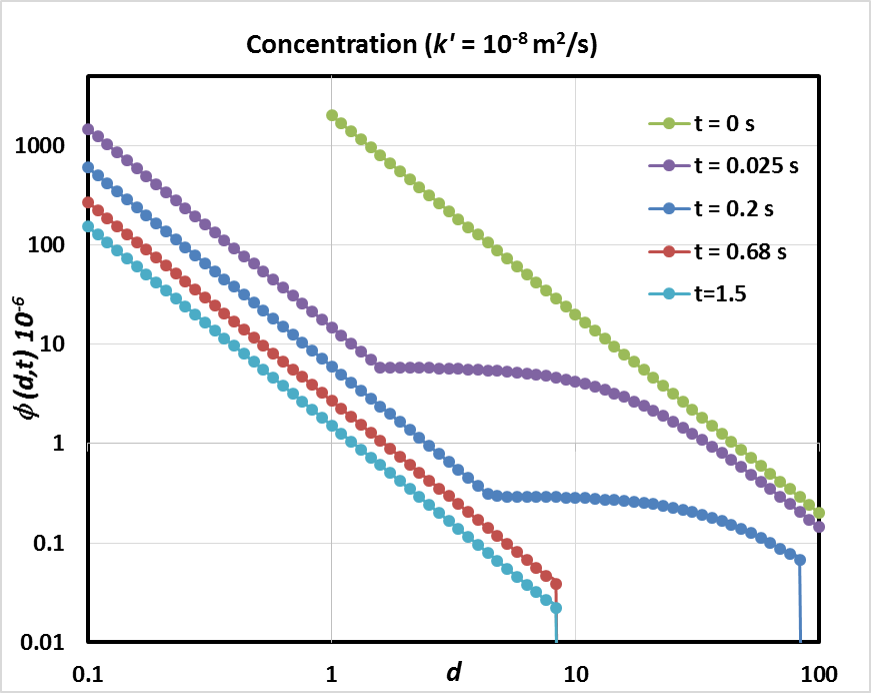}
	\caption{\label{spectra3} Droplet/aerosol concentration evolution as predicted by the analytical model  presented in \eqref{spec1} and \eqref{spec2}.}
\end{figure}

In this section we will demonstrate the efficacy of the simple model presented in \eqref{spec1} and \eqref{spec2} for the prediction of droplet/aerosol concentration. In contrast to the Monte-Carlo approach of the previous subsection, where the evolution of each droplet was accurately integrated, here we will use the analytical prediction along with its simplifying assumptions. The cases considered are identical to those presented in Figure~\ref{spectra2} for $\psi = 0.1\%$ and $k' = 10^{-8} \, m^2/s$. The initial droplet size distributions considered are again log-normal and Pareto distributions. In this case, however, we underline that the quantity of importance in airborne transmission is not the total number of droplet residues, but rather their concentration in the proximity of a susceptible host. Accordingly, we plot in Figure~\ref{spectra3} airborne droplet and residue concentration (per liter) of volume as a function of droplet size. Here the area under the curve between any two diameter yields the number of droplets within this size range per liter of volume within the cloud. At the early times of $t=0.025$ and $0.2 \, s$, we see that larger droplets above a certain size have fallen out of the cloud, while droplet residues smaller than $d_{evap}$ have fully evaporated and their distribution is a rigidly-shifted version of the original distribution. The distribution of intermediate size airborne droplets reflects the fact that they are still undergoing evaporation. Unlike in Figure~\ref{spectra2}, the concentration continues to fall even after $t_{lim} \simeq 0.68 \, s$ when the number and size of droplets within the cloud have reached their limiting value. This is simply due to the fact that the volume of the puff continues to increase and this continuously dilutes the aerosol concentration. Most importantly, the results of the simple model  presented in \eqref{spec1} and \eqref{spec2} are in excellent agreement with those obtained from Monte-Carlo simulation. The increasing size of the contaminated cloud with time can be predicted with \eqref{Qt} and the centroid is given by the scaling law \eqref{eq_Conly}.

\section{Conclusions and Future Perspectives}
The primary goal of this paper is to provide a unified theoretical framework that accounts for all the physical processes of importance, from the ejection of droplets by breathing, talking, coughing and sneezing to the inhalation of resulting aerosols by the receiving host. These processes include: (i) forward advection of the exhaled  droplets with the puff of air initially ejected; (ii) growth of the puff by entrainment of ambient air and its deceleration due to drag; (iii) gravitational settling of  some of the droplets out of the puff; (iv) {modeling of droplets evaporation, assuming that the $d^2$-law prevails}; (v) presence of non-volatile {compounds which form the droplet residue left behind after evaporation; (vi) late-time dispersal of the droplet residue-laden cloud due to ambient air turbulent dispersion. }

Despite the complex nature of the physical processes involved, the theoretical framework results in a simple model for the airborne droplet and residue concentration within the cloud as a function of droplet diameter and time, which is summarized in equations \eqref{spec1}, \eqref{spec2} and \eqref{Qt}. This framework can be used to calculate the concentration of virus-laden residues at the location of any receiving host as a function of time. As additional processes, the paper also considers (vii) efficiency of aspiration of the droplet residues by the receiving host; and (viii) effectiveness of different kinds of masks in filtering the residues of varying size. 

It must be emphasized that the theoretical framework has been designed to be simple and therefore involves a number of simplifying assumptions. Hence, it must be considered as the starting point. By relaxing the approximations and by adding additional physical processes of relevance, more complex theoretical models can be developed. One of the primary advantages of such a simple theoretical framework is that varying scenarios can be considered quite easily: these different scenarios include varying initial puff volume, puff velocity, number of droplets ejected, their size distribution, non-volatile content, ambient temperature, humidity, and ambient turbulence. 

The present theoretical framework can be, and perhaps must be, improved in several significant ways in order for it to become an important tool for reliable prediction of transmission. (i) Accurate quantification of the initially ejected droplets still remains a major challenge. Further high-quality experimental measurements and high-fidelity simulations are required, especially mimicking the actual processes of talking, coughing and sneezing, to fully understand the entire range of droplet sizes produced during the exhalation process. (ii) As demonstrated above, the rate at which an ejected droplet evaporates plays an important role in determining how fast they reach their fully-evaporated state. It is thus important to calculate more precisely the evaporation rate of non-volatile-containing realistic droplets resulting from human exhalation. The precise value of evaporation rate may not be important when droplets evaporate fast, since all droplets remaining within the puff would have completed their evaporation. But under slow evaporation conditions, accurate evaluation of evaporation  {is important}. (iii) The assumption of uniform spatial distribution of droplets within the puff and later within the dispersing cloud is a serious approximation \cite{sbrizzai}. {The intermittency of turbulence within the initial puff and later within the droplet cloud is important to understand and couple with the evaporation dynamics of the droplets. In addition to the role of intermittency, }even the mean concentration of airborne droplets and residues may decay from the center to the outer periphery of the puff/cloud. Characterization of this inhomogeneous distribution will improve the predictive capability of the model. (iv) The presence of significant ambient mean flow and turbulence either from indoor ventilation or outdoor cross-flow will greatly influence the dispersion of the virus-laden droplets. But accounting for their effects can be challenging even in experimental and computational approaches. Detailed experiments and highly-resolved simulations of specific scenarios should be pursued. But it will not be possible to cover all possible scenarios with such an approach. A simpler approach where the above theoretical framework can be extended to include additional models such as random flight model (similar to those pursued in the calculation of atmospheric dispersion of pollutants \cite{hann}) may be promising approaches.

\section{Acknowledgments}
SB acknowledges support from the Office of Naval Research (ONR) as part of the Multidisciplinary University Research Initiatives (MURI) Program, under grant number N00014-16-1-2617 and from the UF Informatics Institute. SZ wishes to thank the French ANR for its support through its flash Covid-19 program - NANODROP grant, the ERS Advanced Grant TRUFLOW and the PRACE network for its Covid-19  Fast Track Call grant NANODROP on the Irene TGCC. AS acknowledges funding from the PRIN project Advanced computations and experiments in turbulent multiphase flow (Project No. 2017RSH3JY). GA acknowledges support through the NSF Grant No. CBET 2029548 and the Clarkson IGNITE Fellowship. LB acknowledges support from the Smith Family
Foundation, the Massachusetts Institute of Technology (MIT) Policy Lab, the MIT Reed Fund, and the Esther and Harold E. Edgerton Career Development chair at MIT.

%\include{Attempt_july1}

%\bibliographystyle{aiaa}
%\bibliography{biblio_stephane}

%\include{covid_19_references}

\end{document}